\documentclass[aps,prd,twocolumn,fleqn,superscriptaddress]{revtex4}
\usepackage[utf8]{inputenc}
\usepackage{graphicx,color,natbib}
\usepackage{amsmath,amssymb,amsfonts}        
\newcommand{\bse}{\begin{subequations}}
\newcommand{\ese}{\end{subequations}}
\newcommand{\be}{\begin{equation}}
\newcommand{\ee}{\end{equation}}
\newcommand{\bea}{\begin{eqnarray}}
\newcommand{\eea}{\end{eqnarray}}
\newcommand{\ba}{\begin{array}}
\newcommand{\ea}{\end{array}}

\input amssym.def
\input amssym.tex

\usepackage[colorlinks=true, linkcolor=blue, bookmarks=true]{hyperref}
\begin{document}
\title{Meson Potential Energy in a Non-Conformal Holographic Model}
\author{M. Asadi\footnote{$\rm{m}_{-}$asadi@ipm.ir}}
\affiliation{School of Particles and Accelerators, Institute for Research in Fundamental Sciences (IPM),
P.O.Box 19395-5746, Tehran, Iran}
\author{Ali Hajilou\footnote{hajilou@ipm.ir}}
\affiliation{School of Particles and Accelerators, Institute for Research in Fundamental Sciences (IPM),
P.O.Box 19395-5746, Tehran, Iran}
\affiliation{Department of Physics, Shahid Beheshti University G.C., Evin, Tehran 19839, Iran}
\begin{abstract}
We study the meson potential energy in a non-conformal model at both zero and finite temperature via gauge/gravity duality. This model consists of five-dimensional Einstein gravity coupled to a scalar field with a non-trivial potential. Interestingly, at both zero and finite temperature we find that the relative meson potential energy can be considered as a measure of non-conformality of the theory. At zero temperature we show that parameters of the Cornell potential, i.e. Coulomb strength parameter $\kappa$ and constant $C$ depends on the energy scale $\Lambda$ that breaks conformal symmetry and the difference between the number of degrees of freedom of UV and IR fixed points $\Delta N$ while QCD string tension $\sigma_s$ just depends on the $\Lambda$. At finite temperature we see that there is a melting length $l_{m\ell}$ where beyond that the meson dissociates in the plasma and by increasing $\Lambda$ the value of $l_{m\ell}$ increases while its value decreases by increasing the temperature.
\end{abstract}
\maketitle


\section{Introduction}
The Anti de-Sitter/Conformal Field Theory (AdS/CFT) correspondence or more generally gauge/gravity duality provides a wide range of domain to study strongly coupled quantum field theories whose duals are the gravitational theories in one higher dimension \cite{Maldacena:1997re,Witten:1998qj,Gubser:1998bc,CasalderreySolana:2011us,Ammon:2015wua}.
In fact, this correspondence is a strong-weak duality that is a novel approach to attack the problems of strongly coupled field theories using the classical gravity dual which is a weakly coupled theory.
Accordingly, parameters, fields, and different phenomena in the gauge theory are translated into equivalents on the gravity side. Therefore, many different questions in the strongly coupled gauge theory can be set up and investigated in dual gravity.
This conjectured duality used to explore applications in the strongly coupled region of Quantum Chromodynamics (QCD), i.e. holographic QCD \cite{Kruczenski:2003be,Kruczenski:2003uq,Kobayashi:2006sb,Babington:2003vm,Sakai:2004cn,Sakai:2005yt,Erlich:2005qh,Karch:2006pv,Li:2011hp,DeWolfe:2010he,DeWolfe:2011ts,Hajilou:2021wmz,Amiri-Sharifi:2016uso,Ali-Akbari:2015bha,Bohra:2020qom,Dudal:2021jav,AliAkbari:2012vt,Ali-Akbari:2013txa,Fang:2015ytf,Callebaut:2011ab,Arefeva:2020vae,Braga:2017bml,Abt:2019tas,Nakas:2020hyo}.
 
One of the important strongly coupled systems is Quark-Gluon Plasma (QGP) produced at Relativistic
Heavy Ion Collider (RHIC) and Large Hadron Collider (LHC) by colliding heavy nuclei at a relativistic speed \cite{Shuryak:2003xe,Shuryak:2004cy}. The potential energy between a quark and an anti-quark living in the plasma is very useful for investigating the strong interactions and quark confinement.
One of the most important non-local and gauge invariant observables in gauge theories is Wilson loop. In quantum field theory, utilizing this gauge invariant and non-local quantity one can obtain the potential energy \cite{Schwartz:2014sze}.
The expectation value of Wilson loop has much power to tackle the interesting questions in QCD such as confinement \cite{Wilson:1974sk}.
Concisely, our approach to calculate the static potential energy of the heavy quark-antiquark pair is to compute the expectation value of a rectangular Wilson loop in the strongly coupled plasma. 
 It is important to note that extracting the meson potential energy from Wilson operator in holography is only valid in the heavy quark limit. For more details refer to \cite{CasalderreySolana:2011us}.
 The holographic dual of the rectangular Wilson loop is given by a classical open string suspended from two points in the boundary of the gravity theory and hanging down in the gravity bulk. In fact, Wilson loop was used to calculate the potential energy of the heavy quark-antiquark for the first time \cite{Maldacena:1998im}. Using this idea, the meson potential energy has been extensively investigated in static backgrounds \cite{Brandhuber:1998bs,Andreev:2006ct,Yang:2015aia} and also is generalized to time-dependent backgrounds \cite{Ali-Akbari:2015ooa,Hajilou:2017sxf,Ishii:2014paa,Hajilou:2018dcb}.

Gauge/gravity duality covers the conformal theories as well as non-conformal ones. It is very interesting to study the effect of non-conformality on the physical quantities. For instance,  the equilibration rates of strongly coupled non-conformal quark-gluon plasmas is studied in \cite{Buchel:2015saa} and the behavior of the potential energy and holographic subregion complexity corresponding to a probe meson in a non-conformal background is investigated in \cite{Lezgi:2020bkc}.
Also, by considering the non-conformal plasma the behavior of the lowest nonhydrodynamic modes in a class of holographic models is investigated in \cite{Janik:2015waa} and the dependence of the approach to thermal equilibrium of strongly coupled non-conformal plasmas is analytically determined in \cite{Gursoy:2015nza}. In addition, sound wave propagation in strongly coupled non-conformal gauge theory plasma
is investigated in \cite{Benincasa:2005iv}.

In this paper in order to study the potential energy of heavy quark-antiquark in a non-conformal model in a holographic set up we consider a five-dimensional Einstein gravity coupled to a scalar field with a non-trivial potential which is introduced and studied in \cite{Attems:2016ugt}. In this model, the corresponding gauge theory is a non-conformal field theory that has two conformal fixed points at ultraviolet (UV) and at infrared (IR). In the gravity side, these solutions are asymptotically AdS$_5$ in the IR and UV limits with different radii. Thermodynamics transport and relaxation properties of this non-conformal background are studied in \cite{Attems:2016ugt}. Also this background utilized for studying entanglement entropy \cite{Rahimi:2016bbv}, mutual and tripartite information \cite{Ali-Akbari:2019zkf} and subregion complexity \cite{Asadi:2020gzl}.

The remainder of this paper is organized as follows. In Section \ref{gr1} we review on the non-conformal background and its holographic dual in the gauge theory side at zero and finite temperature. In Section \ref{gr2} we introduce the rectangular Wilson loop and we then calculate the potential energy between a quark and an antiquark in the plasma. In Section \ref{gr3} we explain our numerical results and describe how parameters of the gauge theory and  in particular the energy scale can affect the meson potential energy. Finally, in Section \ref{gr4} we discuss our main results.

\section{Review on the backgrounds} \label{gr1}
We are interested in studying the meson potential energy of a holographic five-dimensional model whose dual four-dimensional gauge theory is not conformal \cite{Attems:2016ugt}.
\subsection{Zero Temperature}
The non-conformal holographic model which we consider here consists of five-dimensional dimensional Einstein gravity coupled to a scalar field with non-trivial potential. The action of this model is given by:
\be \label{action1}
S=\frac{2}{G_5^2} \int d^5x \sqrt{-g} \left[\frac{1}{4} {\cal{R}} - \frac{1}{2} (\nabla \phi)^2 -V(\phi) \right] \,,
\ee%
where $G_5$ is five-dimensional Newton constant and ${\cal{R}}$ is the Ricci scalar. $\phi$ and $V(\phi)$ are also the scalar field and its potential, respectively.
The non-trivial potential is chosen to be negative and has the following form 
\be \label{pot1}
L^2 V(\phi)=-3-\frac{3}{2} {\phi}^2 -\frac{1}{3}{\phi}^4+\left( \frac{1}{3{\phi}_M^2}+\frac{1}{2{\phi}_M^4}\right){\phi}^6 -\frac{1}{12{\phi}_M^4} {\phi}^8 .
\ee%
This potential possesses a maximum at $\phi=0$ and a minimum at $\phi={\phi}_M >0$, where ${\phi}_M$ is the model parameter. On the bulk side each of these extrema  corresponds to an AdS$_5$ solution with different radii. It is easily seen that the radii of these solutions take the form:
\begin{equation} \label{uvir}
 L= \sqrt{-~\frac{3}{V(\phi )}} = 
    \begin{cases}
      L_{UV}=L ~~~~~~~~~~ \phi=0 ~,\\
                                    \\
      L_{IR}= \frac{L}{1+\frac{{\phi}_M^2}{6}} ~~~~~\phi={\phi}_M ~.
    \end{cases}       
\end{equation}%
From the above equation it is clearly seen that $L_{IR}<L_{UV}$. On the gauge theory side each of these solutions are dual to the UV fixed point at $\phi=0$ and the IR fixed point at $\phi={\phi}_M$.
According to gauge/gravity dictionary, each of these fixed points has a number of degrees of freedom  $N^2$ proportional to $\frac{L^3}{G_5^2}$.
We can calculate the smaller number of degrees of freedom live in the IR fixed point, as indicated by the fact that $L_{IR}<L_{UV}$.
Furthermore, the difference in degrees of freedom between the UV and IR fixed points will be increased if one increases ${\phi}_M$.

The vacuum solution to the Einstein equations for arbitrary ${\phi}_M$ can be parameterized as:
\be \label{metric1}
ds^2=e^{2A(r)}\left( -dt^2 +d{\vec{x}}^2 \right) +dr^2 ~,
\ee%
where
\be  \begin{split} %
\label{metric2}
 e^{2A(r)}=&\frac{{\Lambda}^2 L^2}{{\phi}^2}\left(1-\frac{{\phi}^2}{{\phi}_M^2}\right)^{\frac{{\phi}_M^2}{6}+1} e^{\frac{-{\phi}^2}{6}} ~ , \cr
{\phi}(r)=&\frac{ \Lambda L ~ e^{\frac{-r}{L}}}{\sqrt{1+\frac{{\Lambda}^2 L^2}{{\phi}_M^2}~e^{\frac{-2r}{L}}} } ~,
 \end{split}
 \ee%
 Where $\Lambda$  is the energy scale which breaks conformal symmetry explicitly. 
By applying successive change of coordinates in the holographic direction as \cite{Attems:2016ugt}:
\be \label{coord} \begin{split} %
 u=& L e^{-r/L} ~ , \cr
   z(u)=& \int_0^u du ~ \frac{L}{u} ~ e^{-A(r)}    ~ ,
\end{split} \ee %
we obtain the form of the metric:
\be \label{metric3}
ds^2=\frac{L_{eff}(z)^2}{z^2}\left( -dt^2 +d{\vec{x}}^2 +dz^2\right)  \,,
\ee%
where $L_{eff}=z e^A$ is a non-trivial function of $z$ such that $L_{eff}(0)=L$ and $L_{eff}(\infty)=L_{IR}$.
\subsection{Finite Temperature}
Here, we review the thermal physics of the holographic non-conformal model described in the previous section. The background solution in Eddington-Finkelstein coordinate is \cite{Attems:2016ugt}:
\begin{figure*}[ht]
\begin{center}
\includegraphics[width=84mm]{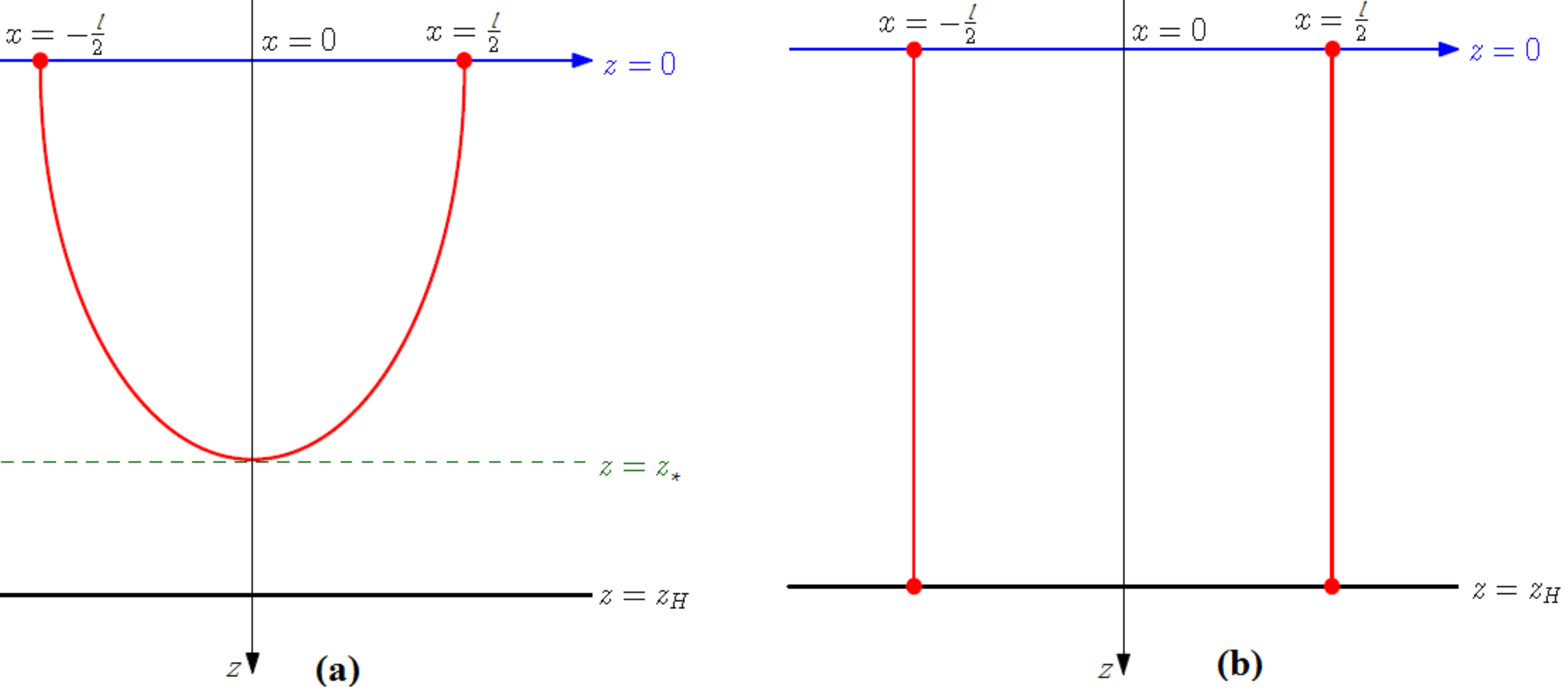}
\hspace{5mm}
\includegraphics[width=84mm]{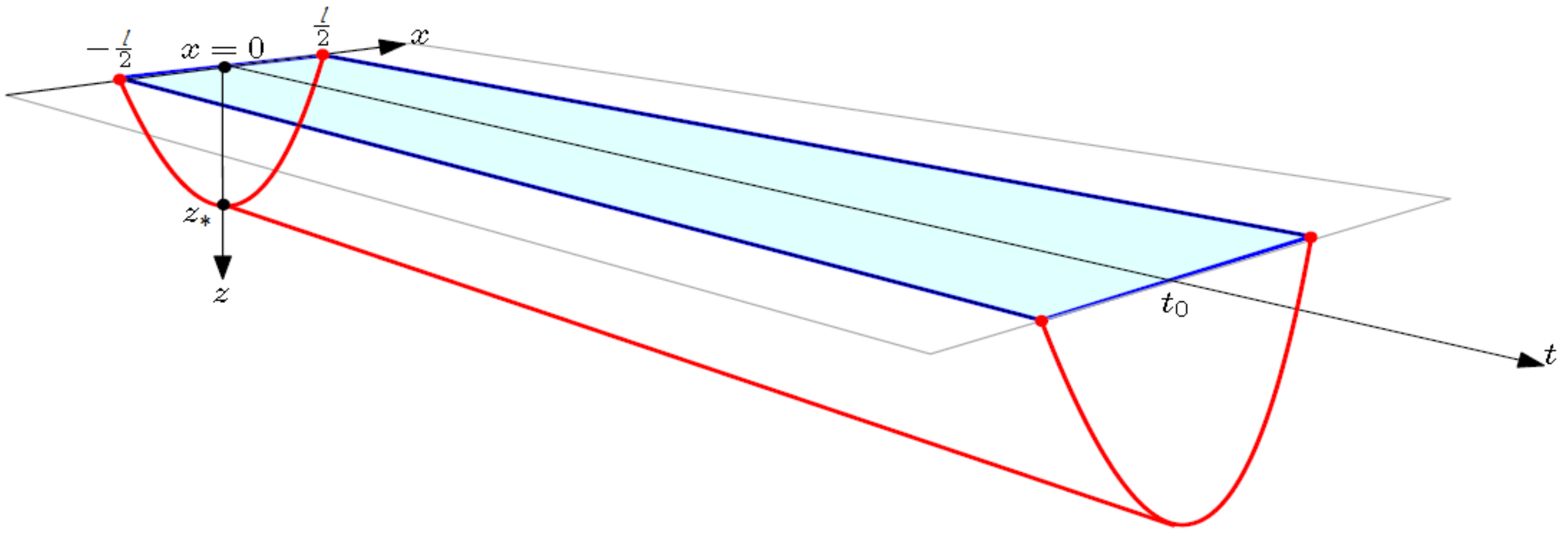}
\caption{Left: Two different configurations of open string. (a): The connected configuration: The $\cup$-shape open string  configuration which connects a quark and an anti-quark on the boundary at $z=0$ and reaches its maximum depth in the bulk gravity at $z=z_\ast$. The coordinate $z$ is the holographic direction in the bulk gravity and $z_\ast$ is the turning point of the open string. (b): The disconnected configuration: The two straight open strings that one of their endpoints is on the boundary at $z=0$ and the other one is at the horizon $z=z_H$.
Right: Wilson loop as the boundary of two-dimensional string world-sheet. The blue rectangular corresponds to the Wilson loop on the boundary of the gravity theory \cite{Yang:2015aia}.
}\label{f11}
\end{center}
\end{figure*}
\be \label{metric20}
ds^2=e^{2A}\left(-h(\phi)d{\tau}^2 +d{\vec{x}}^2 \right) -2e^{A+B} L ~ d\tau d\phi \,,
\ee%
where $h(\phi)$ vanishes at horizon, i.e.  $h(\phi_H)=0$. Utilizing the above ansatz and solving Einstein's equations obtained from (\ref{action1}), the different coefficients of the metric (\ref{metric20}) are given by \cite{Attems:2016ugt}: 
\be \label{component} \begin{split} %
 A(\phi)=& -\log \left( \frac{\phi}{\Lambda}\right)+ \int_0^\phi d\tilde{\phi} \left(G(\tilde{\phi})+\frac{1}{\tilde{\phi}} \right) ~ , \\
 B(\phi)=& \log (|G(\phi)|)+ \int_0^\phi d\tilde{\phi} ~\frac{2}{3G(\tilde{\phi})} ~ ,\\
 h(\phi)=& - \frac{e^{2B} L^2 \left(4V(\phi)+3G(\phi)V'(\phi)\right)}{3G'(\phi)}~,
\end{split} \ee %
where $G(\phi)$ is a master function defined as $G(\phi)=\frac{d}{d \phi} A(\phi)$ and 
satisfies the following non-linear equation:
  \begin{align} \label{component2}
 \frac{G'(\phi)}{G(\phi)+ \frac{4V(\phi)}{3V'(\phi)}}= &\frac{d}{d\phi}\log \Bigg( \frac{1}{3G(\phi)}-2G(\phi) \nonumber \\
 &+\frac{G'(\phi)}{2G(\phi)}-  \frac{G'(\phi)}{2\left(G(\phi)+ \frac{4V(\phi)}{3V'(\phi)}\right)} \Bigg)     \,.
\end{align}  %
The Hawking temperature of the above solution is:
\be \label{component4}
\frac{T}{\Lambda}=- \frac{ L^2  V(\phi_H)}{3 \pi \phi_H} \exp \left[ \int_0^{\phi_H} d\phi \left(    G(\phi) +\frac{1}{\phi}+ \frac{2}{3 G(\phi)}\right) \right].
\ee%

\section{Wilson Loop and Heavy Quark-Antiquark Potential Energy} \label{gr2}
 To do so, we utilize the Wilson loop operator. The expectation value of this operator on a rectangular loop $\cal C$, whose two sides are $l$ and $\cal{T}$ gives us the potential energy between a static quark and an anti-quark, see figure (\ref{f11}). The left  panel of this figure describes the connected and disconnected configuration of the open string and the right panel illustrates the rectangular Wilson loop as the boundary of two-dimensional string world-sheet. If we suppose ${\cal{T}}\gg l$ which means the world-sheet is translationally invariant along the time direction, the expectation value of Wilson loop is \cite{,CasalderreySolana:2011us}:
\be \label{wilson1}
\langle W({\cal{C}}) \rangle =e^{-i(2 m + V(l)){\cal{T}}},
\ee%
where $m$ is the rest mass of quark (or anti-quark) and $V(l)$ is the potential energy between the pair. 
On the other hand, from gauge/gravity duality we know that the expectation value of Wilson loop in the saddle point approximation is dual with the on-shell action of classical open string that suspended from boundary into the bulk and its endpoints correspond to a quark and an anti-quark on the boundary separated with distance $l$:
\be \label{wilson2}
\langle W({\cal{C}})\rangle =e^{i S({\cal{C}})},
\ee%
where $S({\cal{C}})$ is the Nambu-Goto action:
\be\label{action10}
S_{NG}=\frac{-1}{2 \pi \alpha'} \int d\tau d\sigma \sqrt{- \det ~(\gamma_s)_{ab}} ~ ,
\ee%
where  $\alpha'\equiv {{l}_s}^{2}$ in which ${l}_s$ is the fundamental length of the string. In addition, we use the coordinate ($\tau$ , $\sigma$) to parameterize the two-dimensional world-sheet of the string and $(\gamma_s)_{ab}=(g_s)_{\mu\nu}\frac{\partial X^\mu}{\partial \xi^a} \frac{\partial X^\nu}{\partial \xi^b}$ is the induced metric on the world-sheet. Here, $(g_s)_{\mu\nu}$ is the bulk metric in the string frame and $X^\mu$ ($\xi^a=\tau,\ \sigma$) are bulk (world-sheet) coordinates, respectively.

In order to study the potential energy of the heavy quark-antiquark, it is more convenient to use the string frame metric. One can obtain the string frame metric via the standard method that is transformation of scalar (dilaton) field, i.e. $(g_s)_{\mu\nu}=e^{\sqrt{\frac{2}{3}}  \phi} (g)_{\mu\nu}$ \cite{Bohra:2019ebj,Critelli:2016cvq,He:2013qq,Gursoy:2007cb,Gursoy:2007er}.
For completeness, we mention that we considered here the scalar field $\phi$ as a dilaton, for more details see \cite{Charmousis:2010zz} and references therein.

In order to calculate the Nambu-Goto action, $S({\cal{C}})$, we consider the general form of the background metric as:
\be \label{metric32}
ds^2=-f_1(z)dt^2 + f_2(z) dz^2 + f_3(z) d{\vec{x}}^2
 \,.
\ee%
we parameterize the two-dimensional string world-sheet as $\tau= t$, $\sigma=x_3\equiv x$ and we choose all other bulk coordinates except $z$ and $x$ to be constant.
 Therefore, $z=z(x)$ describes the shape of the string and we set a quark and an anti-quark with distance $l$ at $x=\frac{-l}{2}$ and $x=\frac{l}{2}$, respectively. Therefore, the action (\ref{action10}) reduces to:
\be\label{action11}
S_{NG}=\frac{-\cal{T}}{2 \pi \alpha'} \int_{-\frac{l}{2}}^{\frac{l}{2}} dx \sqrt{f_1(z) f_3(z)+ f_1(z) f_2(z) ~ z'^2 \,} ~ ,
\ee%
where $z'\!=\!dz/dx$. Since, the above Lagrangian does not depend explicitly on $x$, then the corresponding Hamiltonian is a constant of motion.
 Using the condition $z'(x)|_{z=z_\ast}=0$ we get:
\be \label{zprime}
z'(x)=\pm \sqrt{\frac{f_1(z)f_3^{2}(z)}{f_1(z_{\star})f_3(z_{\star})f_2(z)}-\frac{f_3(z)}{f_2(z)} \,} ~ ,
\ee%
where $z_\ast=z(x)|_{x=0}$. Using the equation (\ref{zprime}) we obtain:
\be  \label{zprime2}
\int_{\pm \frac{l}{2}}^0 ~ dx= \mp ~  \int_{\epsilon}^{z_{\star}}dz ~ \frac{1}{\sqrt{\frac{f_1(z)f_3^{2}(z)}{f_1(z_{\star})f_3(z_{\star})f_2(z)}-\frac{f_3(z)}{f_2(z)} \,}}
~,
\ee%
where, $\epsilon$ is the IR regulator in the gravity theory that according to the UV/IR connection corresponds to the UV cut-off in gauge theory side \cite{CasalderreySolana:2011us}.
\begin{figure*}[ht]
\begin{center}
\includegraphics[width=58mm]{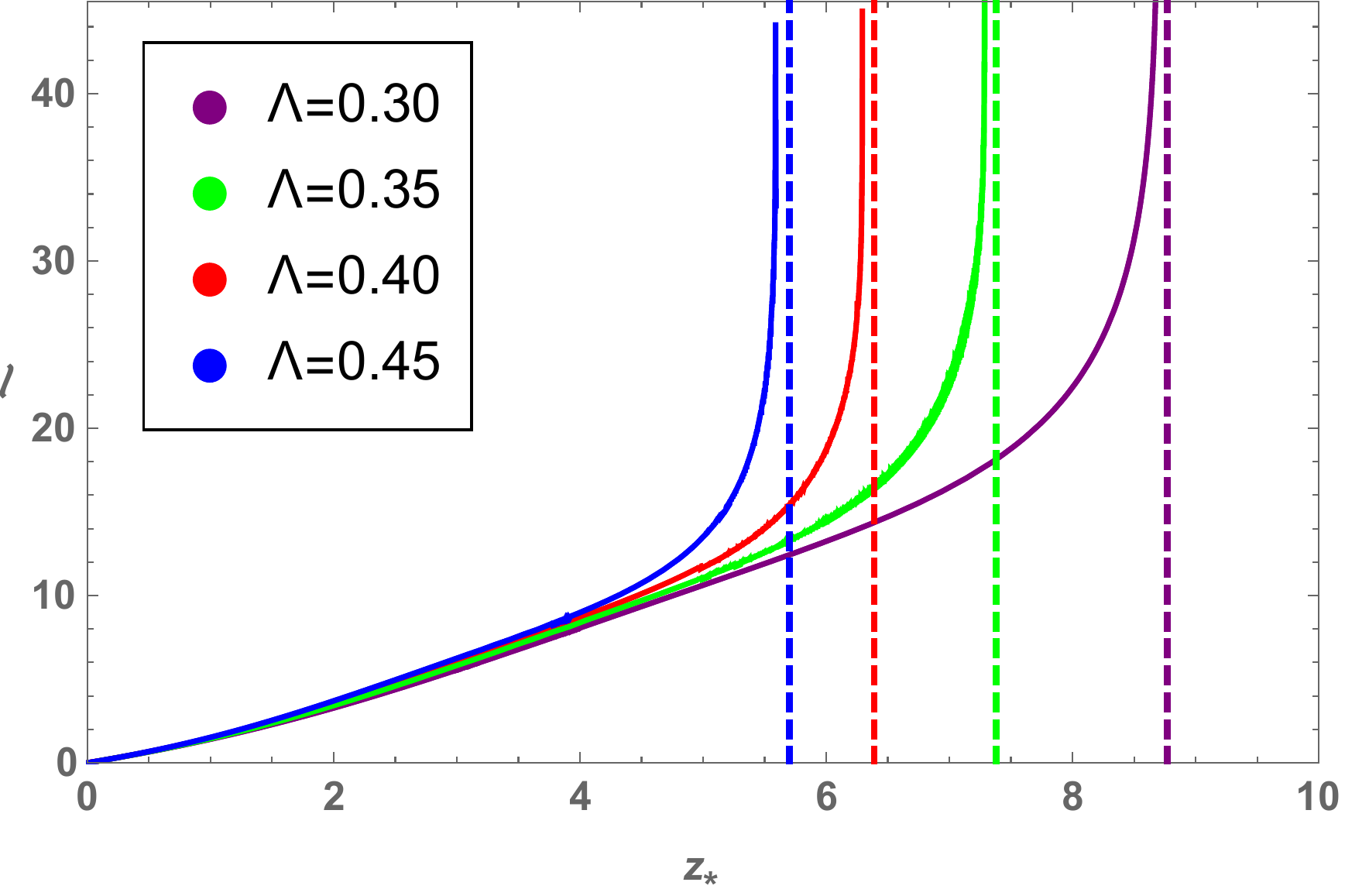}
\includegraphics[width=59mm]{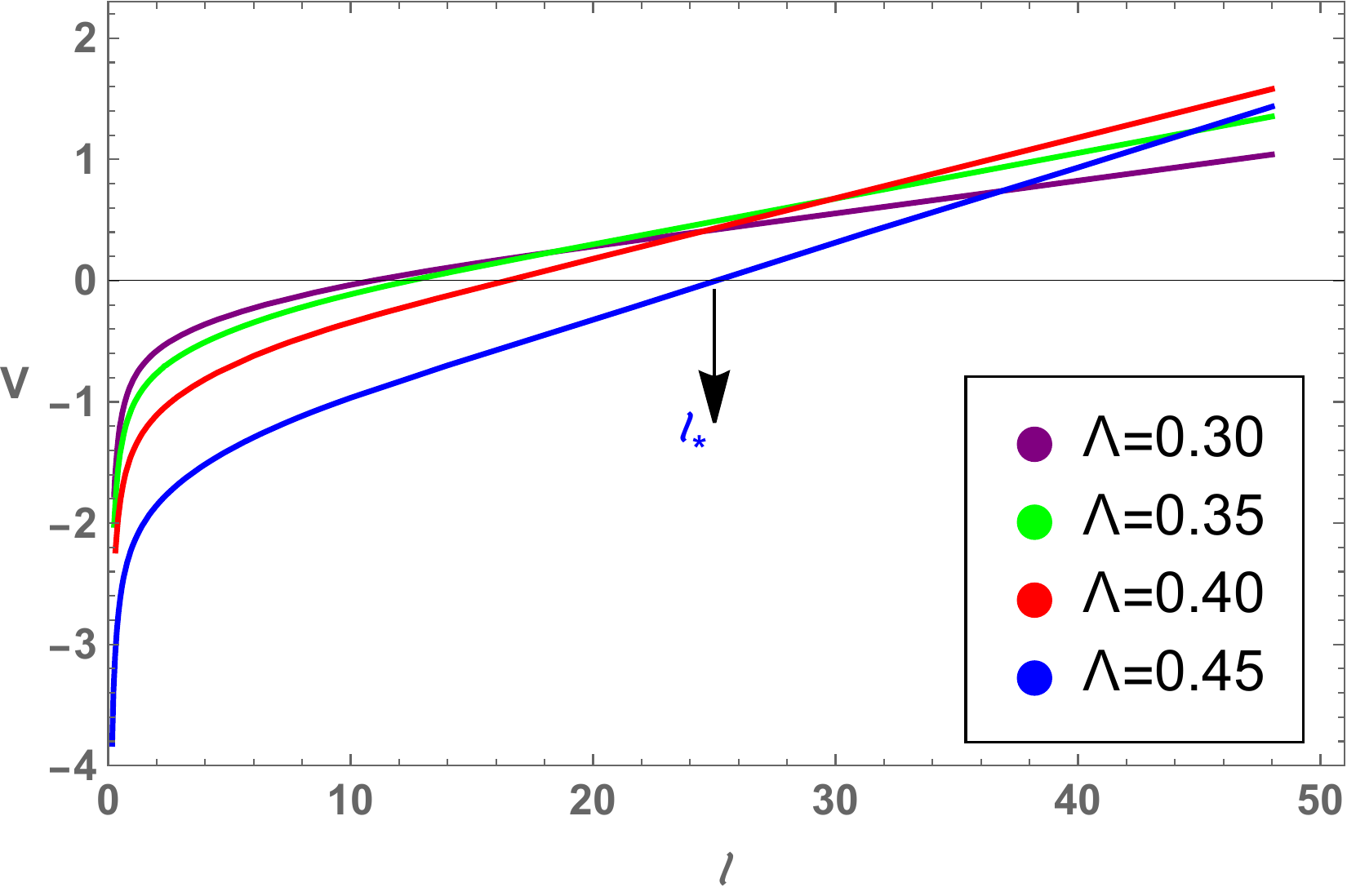}
\includegraphics[width=60mm]{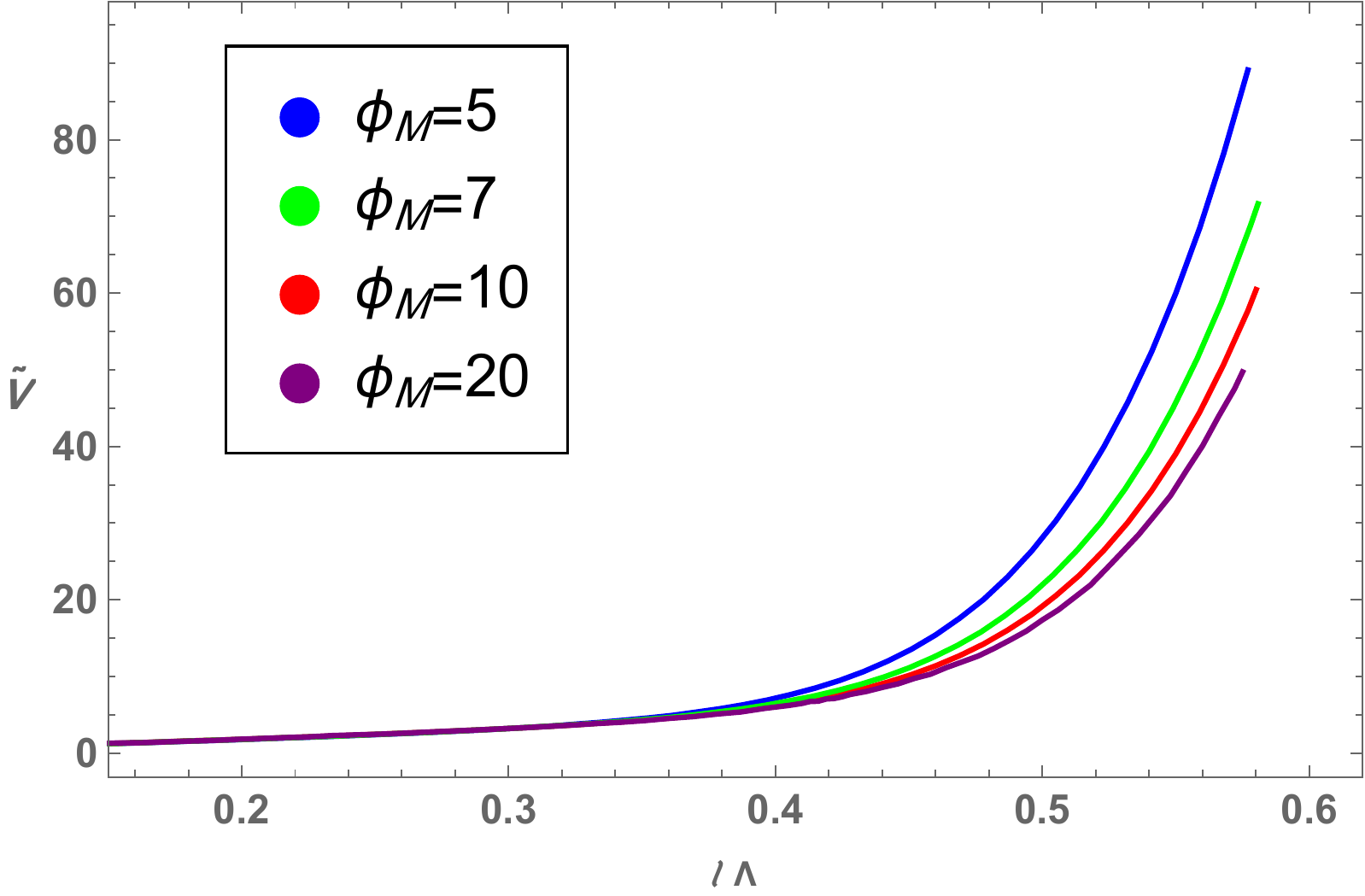}
\caption{ Left: Interquark distance $l$ as a function of $z_\ast$ for different values of the energy scale $\Lambda=0.30$ (Purple), $\Lambda=0.35$ (Green), $\Lambda=0.40$ (Red) and $\Lambda=0.45$ (Blue). We fix the value of model parameter $\phi_M=10$.
Middle: Potential energy of meson as a function of interquark distance $l$ for different values of the energy scale $\Lambda=0.30$ (Purple), $\Lambda=0.35$ (Green), $\Lambda=0.40$ (Red) and $\Lambda=0.45$ (Blue). We fix the value of model parameter $\phi_M=10$.
Right: Relative potential energy of meson $\tilde{V}=\frac{V-V_{AdS}}{V_{AdS}}$ as a function of $l \Lambda$ for different values of the model parameter $\phi_M=5$ (blue), $\phi_M=7$ (Green), $\phi_M=10$ (red) and $\phi_M=20$ (Purple). We fix the interquark distance $l=1$. 
}\label{f1}
\end{center}
\end{figure*}
One can obtain the on-shell action by inserting equations (\ref{zprime}) and (\ref{zprime2}) into equation  (\ref{action10}). In addition, the potential energy of the heavy quark-antiquark pair can be obtained by utilizing equations (\ref{wilson1}) and (\ref{wilson2}). It is important to note that to get a finite value for the potential energy, we need to subtract the rest mass of the quarks from the on-shell action. 
The rest mass of a quark and  an anti-quark is equal to:
\be \label{restmass}
2m =\frac{1}{ \pi \alpha'} \int_{\epsilon}^{zh} dz \sqrt{f_1(z) f_2(z) \,} ~ .
\ee%
Therefore, the potential energy between a quark and an anti-quark is:
 \begin{align} \label{potential10}
V =\frac{1}{ \pi \alpha'} & \Bigg[  \int_{\epsilon}^{z_{\star}} dz \Bigg( \frac{ f_1(z) \sqrt{f_2(z) f_3(z) \,}}{\sqrt{f_1(z) f_3(z)- f_1(z_{\star})f_3(z_{\star}) \,}} \nonumber \\ 
& - \sqrt{f_1(z) f_2(z) \,} \Bigg) 
  -  \int_{z_{\star}}^{zh} dz \sqrt{f_1(z) f_2(z) \,} \Bigg] ~ .
\end{align}
\section{Numerical Results} \label{gr3}
Having set up the ingredients of studying the potential energy of the meson, we are ready to investigate the numerical results for zero and finite temperature cases. Hereafter, in all of our calculations we set $L=1$. 
\subsection{Zero Temperature}
At zero temperature, using the general form of the metric (\ref{metric32}) and the metric (\ref{metric3}) in string frame we have:
\be \label{restmass}
f_1(z)=f_2(z) =f_3(z)= \frac{L_{eff}(z)^2 ~ e^{\sqrt{\frac{2}{3}}  \phi} }{z^2}  ~ .
\ee%
Therefore, one can obtain the potential energy of the heavy quark-antiquark at zero temperature utilizing equations (\ref{potential10}) and (\ref{restmass}).

We plot the behavior of interquark distance $l$ as a function of $z_\ast$ in the left panel of figure (\ref{f1}) for different values of the energy scale $\Lambda=0.30$ (Purple), $\Lambda=0.35$ (Green), $\Lambda=0.40$ (Red) and $\Lambda=0.45$ (Blue) and fixed value of the  model parameter $\phi_M=10$.
According to the left panel, the corresponding potential energy of meson as a function of interquark distance $l$ for different values of the energy scale $\Lambda=0.30$ (Purple), $\Lambda=0.35$ (Green), $\Lambda=0.40$ (Red) and $\Lambda=0.45$ (Blue) and fixed value of the model parameter $\phi_M=10$ is depicted in the middle panel.
 In the right panel, we depict the relative potential energy of meson $\tilde{V}$ as a function of $l \Lambda$ that is defined
 \be \label{relative1}
\tilde{V}=\frac{V-V_{AdS}}{V_{AdS}}  ~,
\ee%
where $V$ and $V_{AdS}$ are  potential energy of meson corresponding to non-conformal and AdS (conformal) background, respectively.
Here, we fixed the value of interquark distance $l=1$ and values of the model parameter $\phi_M=5$ (blue), $\phi_M=7$ (Green), $\phi_M=10$ (red) and $\phi_M=20$ (Purple). 
We list the following results.
\begin{itemize}

\item Left panel
  
It is seen that by increasing $l$, a dynamical "imaginary wall" appears in the bulk where beyond that the connected open string world sheet does not propagate. The physical interpretation is that, we can increase interquark distance $l$ in such a way that the open string that connects a quark and an anti-quark is always in the connected configuration.
 Accordingly, the dual interpretation in the gauge theory side is that a quark and an anti-quark form a bound state such that in all length scales we can investigate the Cornell potential. In other words, we can interpret this case as the confined phase of the theory.
 In addition, the location of the "imaginary wall" shifts to higher values of $z$ by decreasing the energy  scale $\Lambda$, corresponding to a deeper penetration into the bulk.
A similar type of "imaginary wall" is reported before in \cite{Bohra:2019ebj}. For more details see \cite{Yang:2015aia,Dudal:2017max,Arefeva:2018hyo} and references therein.

 \item Middle panel
 
It is seen that in the non-conformal background the meson potential energy increases (becomes more positive) by increasing the interquark distance $l$.
We find that each of the curves in this panel can be fitted with the Cornell potential \cite{Eichten:1978tg,Andreev:2006ct,Bohra:2019ebj,Bruni:2018dqm,Yang:2015aia}:
\be \label{cornel}
V(l)=-\frac{\kappa}{l}+ \sigma_s l + C ~,
\ee%
where, $\kappa$ is a Coulomb strength parameter, $\sigma_s$ is QCD string tension and $C$ is a constant. 
It is clearly seen from (\ref{cornel}) that the Coulomb part of the Cornell potential dominates in the ultraviolet (UV) region and its linear part dominates in the infrared (IR)
region. To put in another way, the Coulomb term, i.e. $-\frac{\kappa}{l}$ dominates for small values of $l$ and the linear term, i.e. $\sigma_s l$ dominates for larger values of $l$.
Another point is that, when we increase the interquark distance $l$ we observe that there is a \textit{specific length}, i.e. $l_{\star}$, where beyond this length the potential energy of meson experiences the linear regime of the Cornell potential, i.e. \textit{confinement}. It is seen that $l_{\star}$ increases by increasing the energy scale $\Lambda$.
Physical description of this behavior is that when we break the conformal symmetry at larger energy scale $\Lambda$ the meson enjoys the linear confinement at larger interquark distance $l$.
Another feature is that in this panel at very small $l$ the two curves coincide. Since, at very small $l$ we work in the UV regime where the conformal symmetry restores and effect of the energy scale $\Lambda$ (that breaks the conformal symmetry) disappears. 
Furthermore, this panel shows that for small interquark distance $l$ (but not very small) where the potential energy is negative, by increasing the energy scale $\Lambda$ the potential energy between the pair decreases (becomes more negative) and the pair gets more bounded. Interestingly, at very large interquark distance $l$ where the linear regime of the Cornell potential is dominated, we also observe that by increasing $\Lambda$ the potential energy increases, i.e. the interaction between the heavy quark and antiquark increases.

 \item Right panel

It is seen that the value of $\tilde{V}$ is always positive ($V>V_{AdS}$) and by increasing $\Lambda$ the value of $\tilde{V}$ increases monotonically. In addition, we observed that when the energy scale $\Lambda$ is small enough, the relative potential energy of meson $\tilde{V}$ are approximately independent of $\phi_M$. This is due to the fact that at small $\Lambda$ the field theory is almost conformal and hence the two curves coincide. 
On the other hand, at large enough $\Lambda$ the value of $\tilde{V}$ depends on $\phi_M$. By increasing $\phi_M$, increasing the difference in degrees of freedom between the UV and the IR fixed points, $\tilde{V}$ decreases and two curves deviate from each other.
Therefore, for small enough $\Lambda$ we can deduce that $\tilde{V}$ is not a good measure to quantify the deviation from the non-conformality while for large enough one we can use $\tilde{V}$ as a measure of the non-conformality of the theory at zero temperature.
 \end{itemize}
\begin{figure*}[ht]
\begin{center}
\includegraphics[width=69mm]{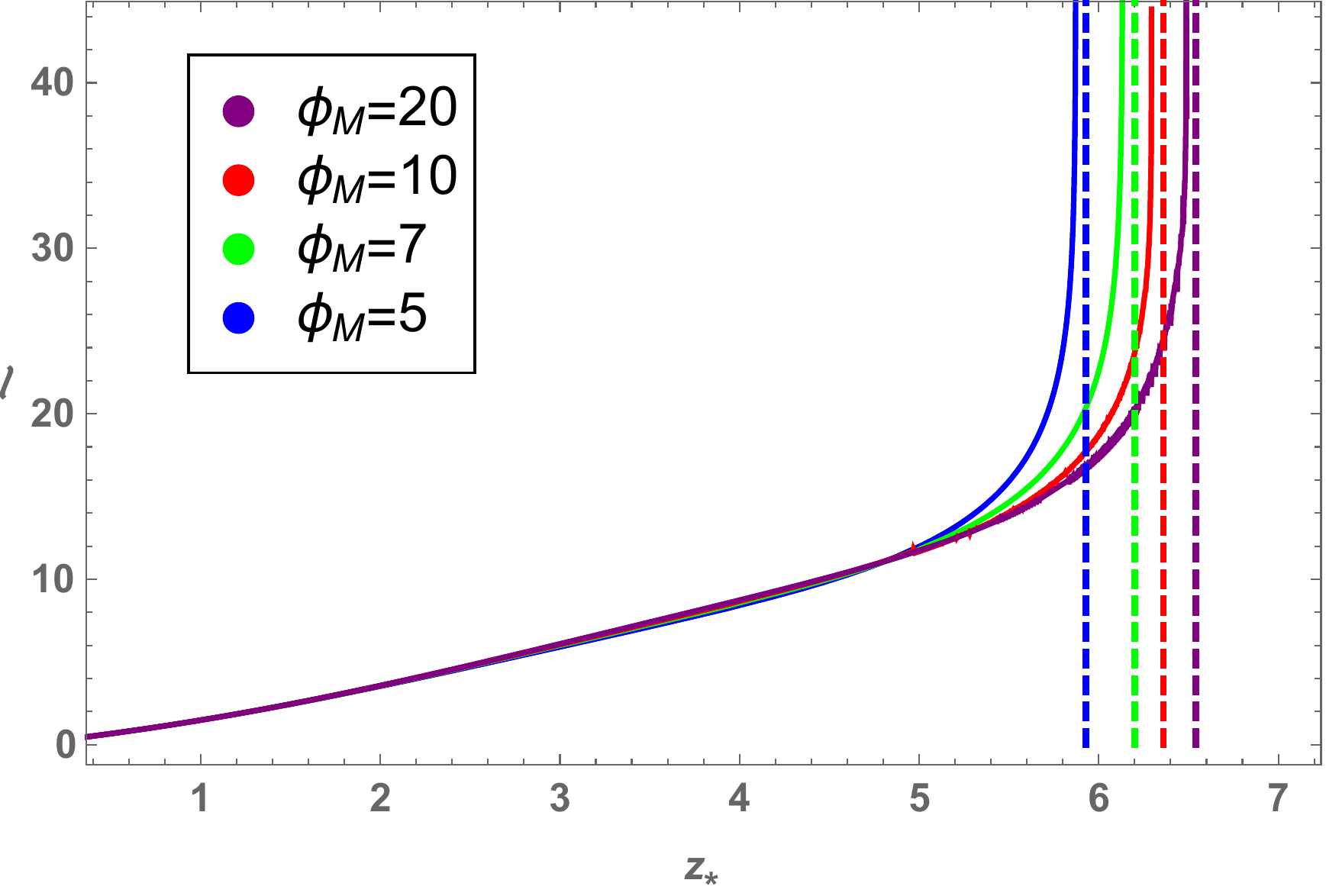}
\hspace{10mm}
\includegraphics[width=71mm]{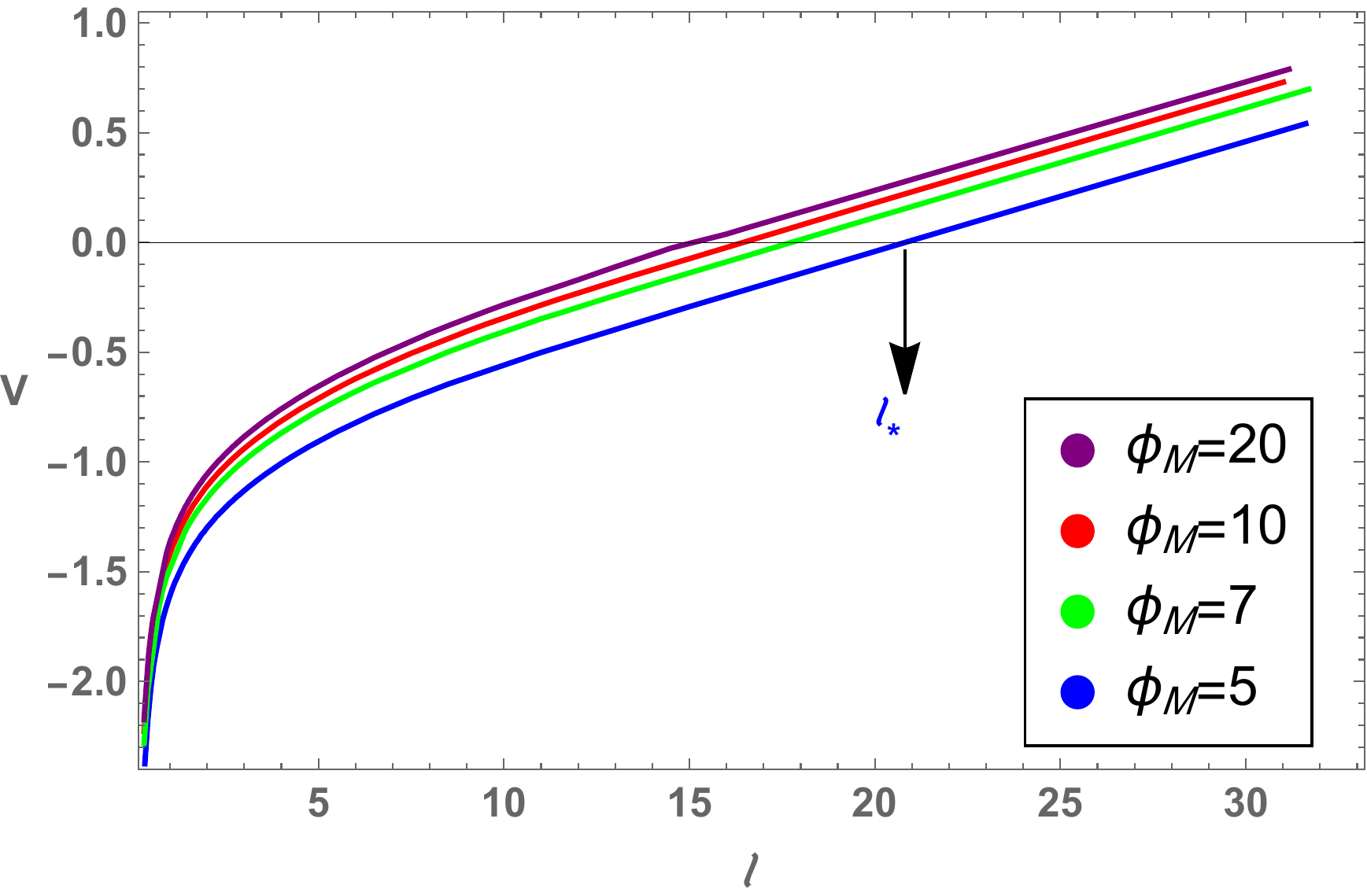}
\caption{ Left: Interquark distance $l$ as a function of $z_\ast$ for different values of the model parameter $\phi_M=5$ (blue), $\phi_M=7$ (green), $\phi_M=10$ (red) and $\phi_M=20$ (purple).  We fix the value of energy scale $\Lambda=0.4$.
Right: Potential energy of meson as a function of interquark distance $l$ for different values of given model parameter $\phi_M=5$ (blue), $\phi_M=7$ (green), $\phi_M=10$ (red) and $\phi_M=20$ (purple). We fix the value of energy scale $\Lambda=0.4$. 
}\label{f22}
\end{center}
\end{figure*}

\begin{figure*}[ht]
\begin{center}
\includegraphics[width=58mm]{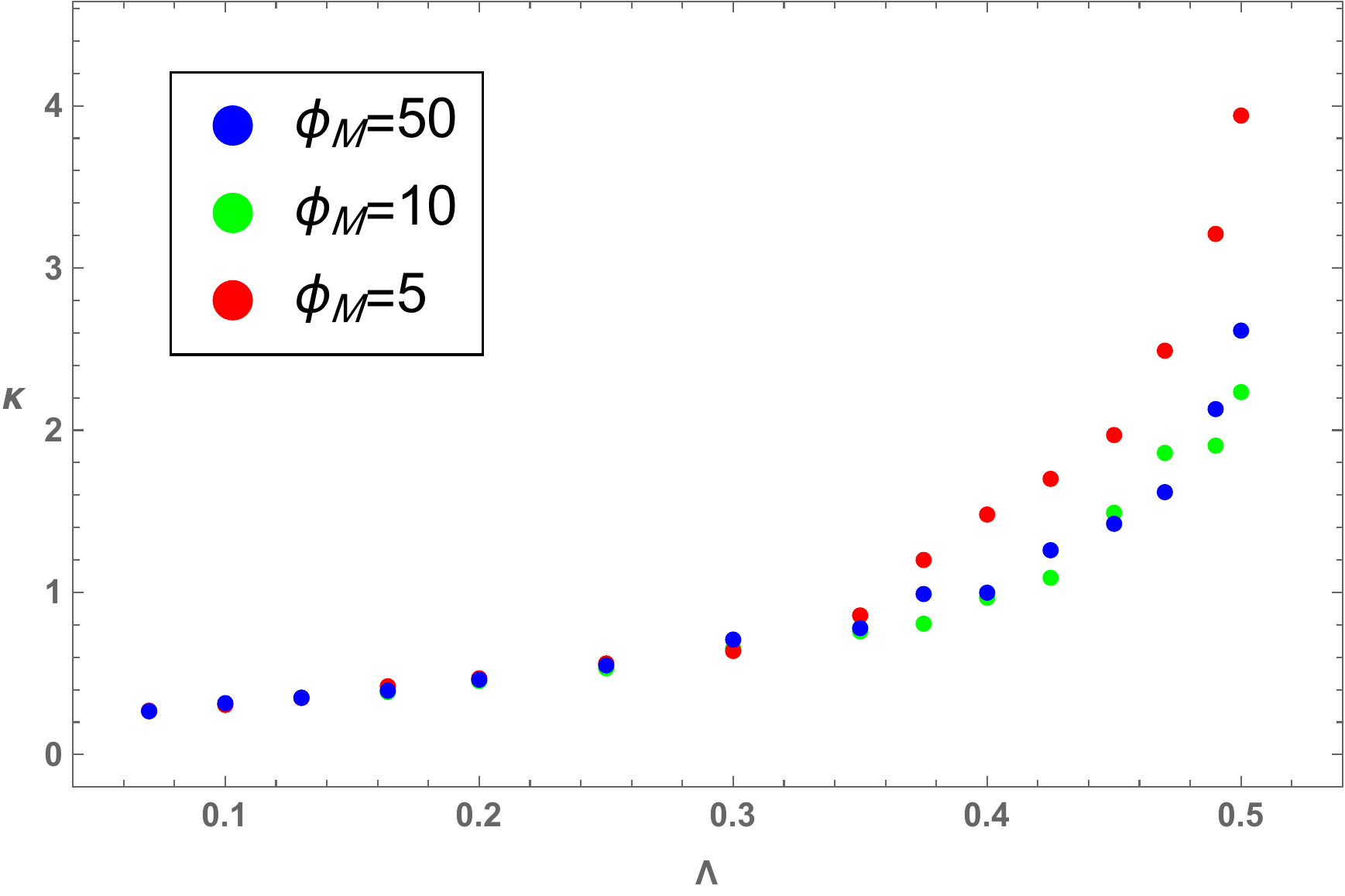}
\includegraphics[width=60mm]{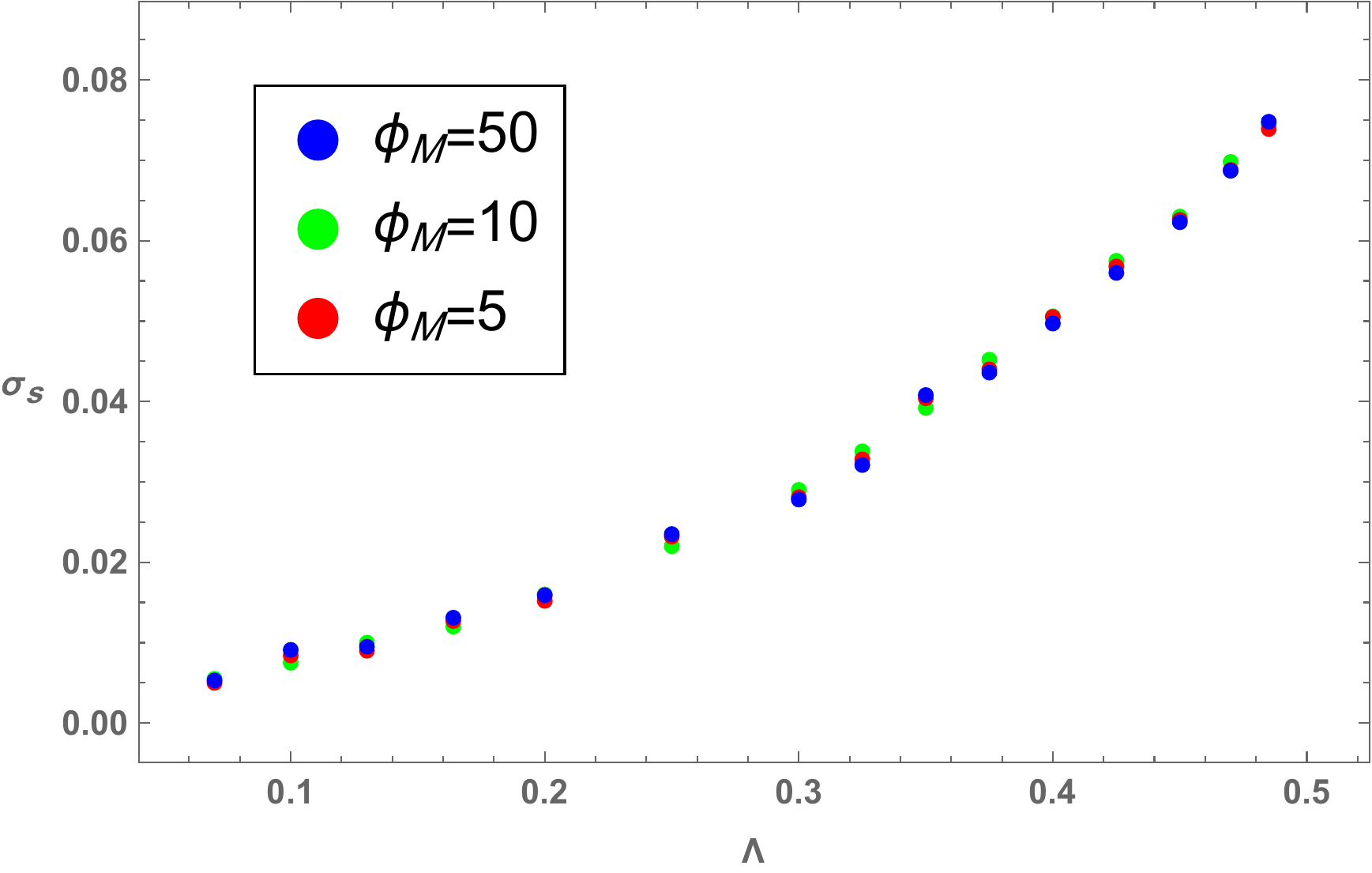}
\includegraphics[width=59mm]{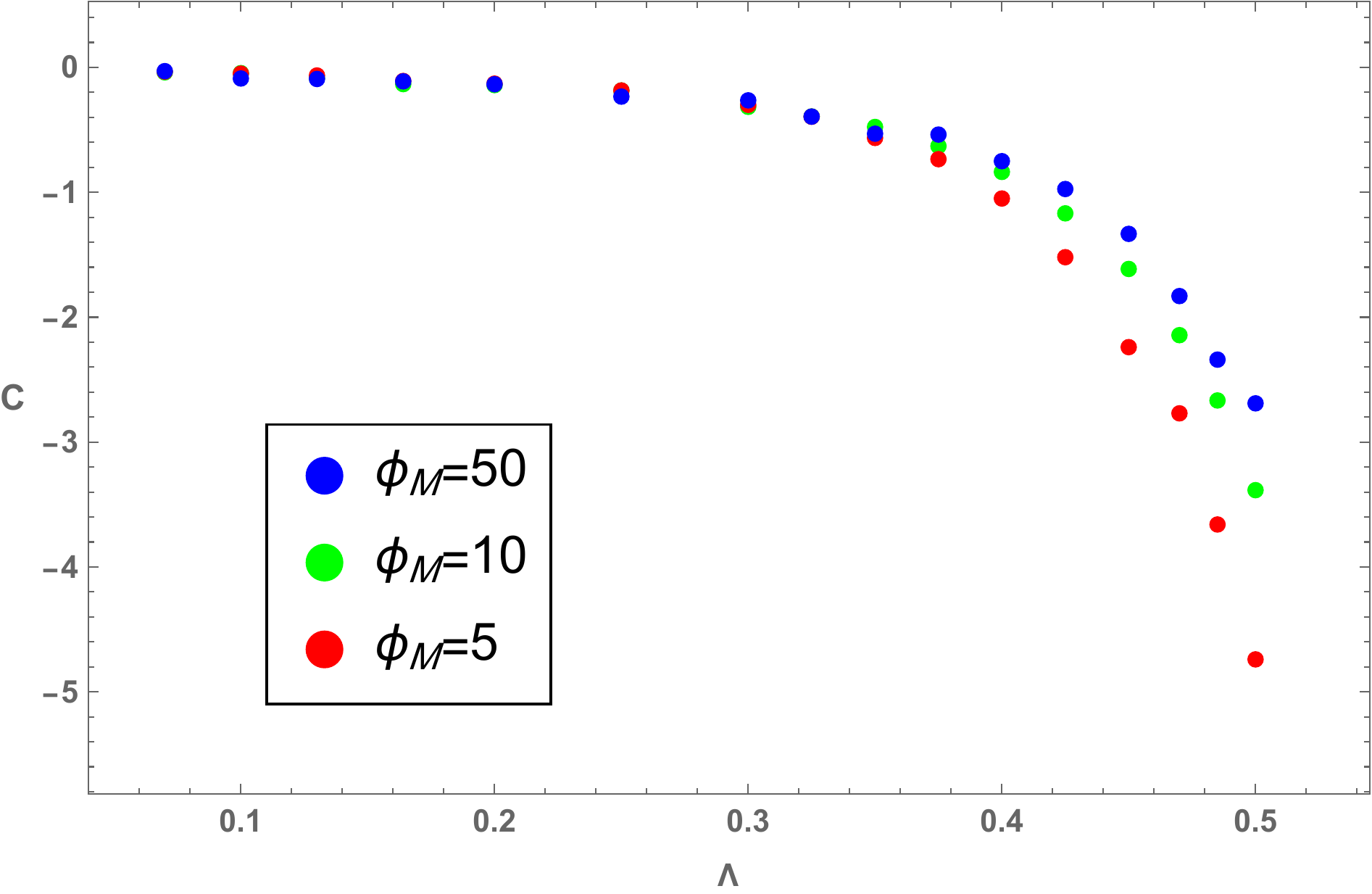}
\caption{Left: Coulomb strength parameter $\kappa$ as a function of energy scale $\Lambda$.
Middle: QCD string tension $\sigma_s$ as a function of the energy scale $\Lambda$.
Right: The constant of the Cornell potential $C$ as a function of energy scale $\Lambda$.
Each panel is plotted for different values of the model parameter $\phi_M=5$ (red), $\phi_M=10$ (green) and $\phi_M=50$ (blue). 
}\label{f2}
\end{center}
\end{figure*}

The dependence of interquark distance $l$ to the $z_\ast$ in the left panel of figure (\ref{f22}) for fixed value of energy scale $\Lambda=0.4$ and  different values of given model parameter $\phi_M=5$ (blue), $\phi_M=7$ (green), $\phi_M=10$ (red) and $\phi_M=20$ (purple).
The corresponding  meson potential energy $V$ related to the left panel is shown as a function of interquark distance $l$ for fixed value of energy scale $\Lambda=0.4$ and  different values of given model parameter $\phi_M=5$ (blue), $\phi_M=7$ (green), $\phi_M=10$ (red) and $\phi_M=20$ (purple).
 We summarize our results in the following.
 \begin{itemize}
\item Left panel

It is seen that by increasing $l$, a dynamical "imaginary wall" appears in the  bulk and the physical interpretation is the same as the left panel of figure (\ref{f1}). In fact, beyond the "imaginary wall" the connected open string world sheet does not propagate. In addition, while we increase the interquark distance $l$ the open string keeps its connected configuration and in the gauge theory side it means a quark-antiquark pair does not dissociate and preserves the bound state. Here, we observe this phenomena for different values of $\phi_M$.
 In addition, in comparison to the left panel of figure (\ref{f1}) the location of the "imaginary wall" shifts to higher values of $z$ by increasing the model parameter $\phi_M$, corresponding to a deeper penetration into the bulk.

\item Right panel

 The effect of the model parameter $\phi_M$ on the meson potential energy is investigated.
We observe that at fixed $\Lambda$, changing $\phi_M$ leads to the change in the value of the potential energy of meson. 
In addition, this panel shows that $l_{\star}$ changes by changing $\phi_M$. It is seen that for $\Lambda=0.4$ for some values of $\phi_M$ by increasing its value the potential energy of meson experiences the linear regime at smaller $l$. However, this result is not general and our numerical results show that for different values of $\Lambda$ and $\phi_M$ we observe different behaviors.
The interesting point is that changing $\phi_M$ does not change the QCD string tension, since all the curves shown present the same slope that is in compatible with the middle panel of the figure (\ref{f2}). 

\end{itemize}
As we said in the previous section, the energy scale $\Lambda$ breaks the conformal symmetry and the model parameter $\phi_M$ affects the difference in degrees of freedom between the UV and IR fixed points. Next, we would like to study the $\Lambda$ and $\phi_M$ dependence of Cornell potential parameters  $\kappa$, $\sigma_s$  and $C$ which are depicted in the left, middle and right panels of the figure (\ref{f2}), respectively. These parameters are plotted as a function of $\Lambda$ for a different values of $\phi_M=5$, $\phi_M=10$ and $\phi_M=50$. Our results are listed as follows.

\begin{itemize}
\item Left panel

 We observe that the Coulomb strength parameter $\kappa$ monotonically increases by increasing $\Lambda$ and physically it means that the value of $\kappa$ depends on the energy scale $\Lambda$. Another point is that at small values of $\Lambda$ the dependence of $\kappa$  on the model parameter $\phi_M$ is not manifest, while for larger values of the energy scale $\Lambda$ we see that the $\kappa$ depends on the $\phi_M$ clearly. Since, $\phi_M$ changes the difference between the degrees of freedom between UV and IR fixed points, i.e. $\Delta N$, we can say that the Coulomb strength parameter $\kappa$ depends on the $\Lambda$ and  $\Delta N$ that is $\kappa (\Lambda,\Delta N)$.

\item Middle panel

To describe our result in this panel let's consider the linear regime of the Cornell potential (\ref{cornel}) where $V \propto \sigma_s l$ and  the QCD string tension $\sigma_s$ can be obtained via $\sigma_s = \frac{d V}{d l}$.
We observe that the QCD string tension $\sigma_s$ monotonically increases when $\Lambda$ increases. To put in another word, by attention to the linear regime of the Cornell potential we see that increasing $\Lambda$, increases the potential energy of the pair, i.e. the pair bound together more stronger. Therefore, We clearly observe  that  increasing  $\Lambda$ leads to the more stable bound state. 
Another important feature is that by changing $\phi_M$ the value of $\sigma_s$ does not change that we mentioned intuitively in the physical description of the right panel of figure (\ref{f22}) . Therefore, we conclude that the $\sigma_s$ just depends on the energy scale $\Lambda$ that is $\sigma_s(\Lambda)$.

\item Right panel

It is seen that the constant $C$ in the Cornell potential (\ref{cornel})
 monotonically decreases by increasing $\Lambda$. Therefore, this means that the value of $C$ depends on the $\Lambda$. Another feature is that at small values of $\Lambda$ the dependence of $C$  on the $\phi_M$ is low, while for larger values of the energy scale $\Lambda$ the $C$ depends on the $\phi_M$ manifestly. Consequently, this indicates that the constant $C$  depends on the $\Lambda$ and  $\Delta N$ that is $C (\Lambda,\Delta N)$.

\end{itemize}
In conclusion, in terms of the above discussion corresponding to the figure (\ref{f2}), in this model the Cornell potential as a function of $l$, $\Lambda$ and $\Delta N$ can be rewritten in the form
\be \label{cornel28}
V(l,\Lambda,\Delta N)=-\frac{\kappa (\Lambda,\Delta N)}{l}+ \sigma_s(\Lambda) l + C(\Lambda,\Delta N) ~,
\ee%
where both $\kappa$ and $C$ are function of $\Lambda$ and $\Delta N$, and $\sigma_s$ is a function of $\Lambda$.
\subsection{Finite Temperature}

At finite temperature, utilizing the metric (\ref{metric20}) and general form of the metric  (\ref{metric32}) in the string frame we have:
\be \label{parts} \begin{split} %
f_1(\phi)=&h(\phi)~e^{2A(\phi)} e^{\sqrt{\frac{2}{3}}  \phi} ~, \\
f_2(\phi)=&\frac{e^{2B(\phi)}  e^{\sqrt{\frac{2}{3}}  \phi}  }{h(\phi)} ~, \\
f_3(\phi)=&e^{2A(\phi)}  e^{\sqrt{\frac{2}{3}}  \phi} ~.
\end{split}
\ee%
Therefore, the potential energy of meson $V$ at finite temperature $T$ can be obtained from equations (\ref{potential10}) and (\ref{parts}).

We plot $V$ as a function of $\frac{\Lambda}{T}$ for fixed values of the model parameter $\phi_M=10$, interquark distance $l=0.2$ and different values of energy scale  $\Lambda=1$ (blue), $\Lambda=0.7$ (green), $\Lambda=0.5$ (red) and $\Lambda=0.2$ (purple) in the left panel of figure (\ref{f6}).
In the middle panel, we plot $V$ as a function of $l{\Lambda}$ for fixed values of given model parameter $\phi_M=10$, the energy scale $\Lambda=0.6$ and different values of the temperature $T=0.0152$ (blue), $T=0.0537$ (red), $T=0.0975$ (green) and  $T=0.1256$ (purple).
In the right panel, we plot $V$ as a function of $l{\Lambda}$ for different values of energy scale $\Lambda=1.2$ (blue), $\Lambda=1$ (red), $\Lambda=0.8$ (green), $\Lambda=0.6$ (purple) and fixed values of the given model parameter and the temperature of the system $\phi_M=10$ and $T=0.0975$, respectively.
Our results are listed as follows.

\begin{itemize}
\item Left panel

It is clearly seen that  the potential energy of meson as a function of $\frac{\Lambda}{T}$ decreases monotonically.
This panel shows that by increasing the temperature $T$ the potential energy between the pair increases and quark and antiquark can not communicate easily, i.e. the meson gets less bounded. 
On the other hand, we see that by increasing $\Lambda$ (at fixed temperature) the meson potential energy decreases and meson gets more bounded. Therefore, this panel shows that the effect of $\Lambda$ on the potential energy of meson $V$ is opposite to the temperature $T$. More illustration on the low and high temperature regions is mentioned in the following.
\begin{figure*}[ht]
\begin{center}
\includegraphics[width=59mm]{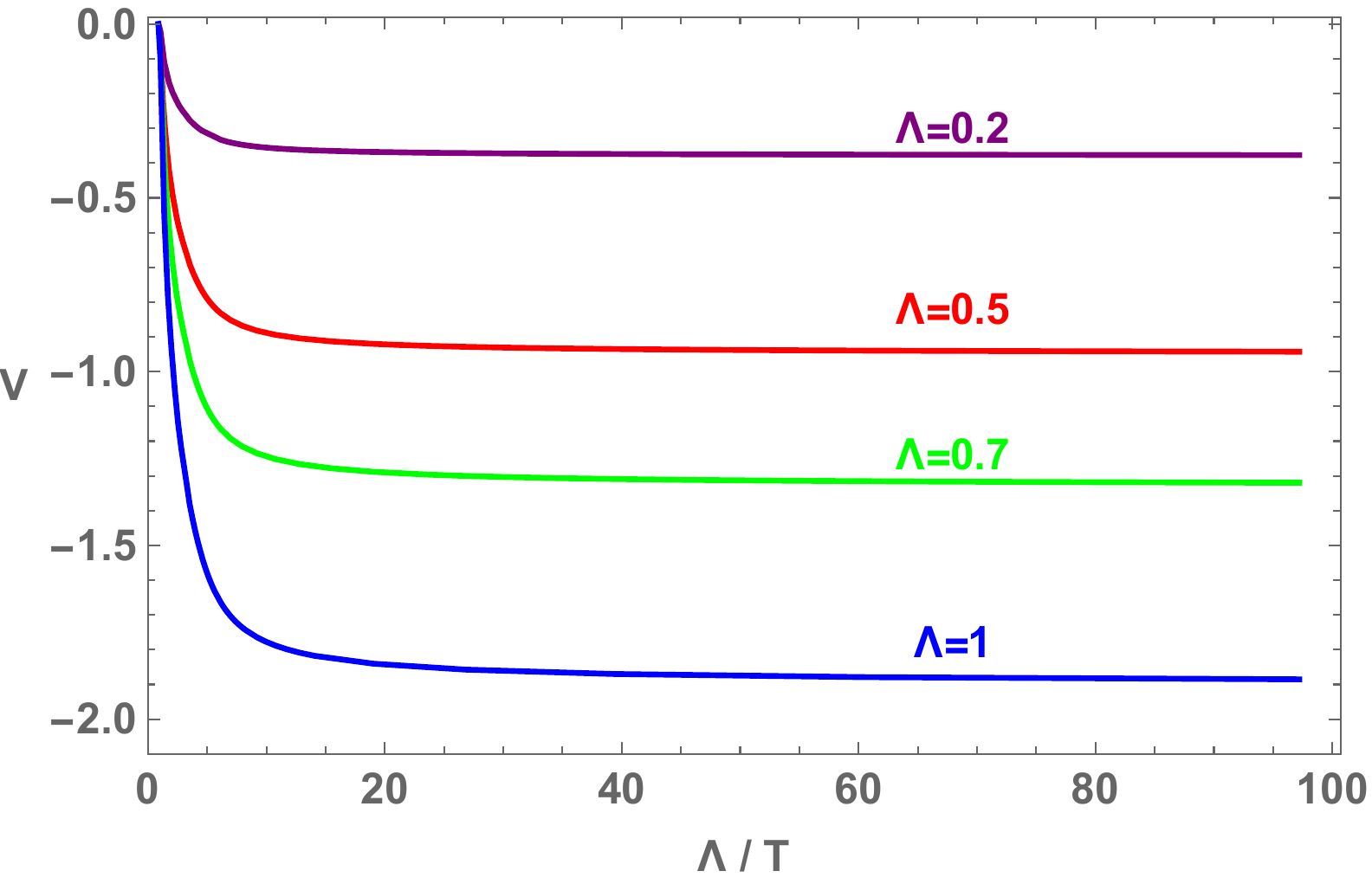}
\includegraphics[width=59mm]{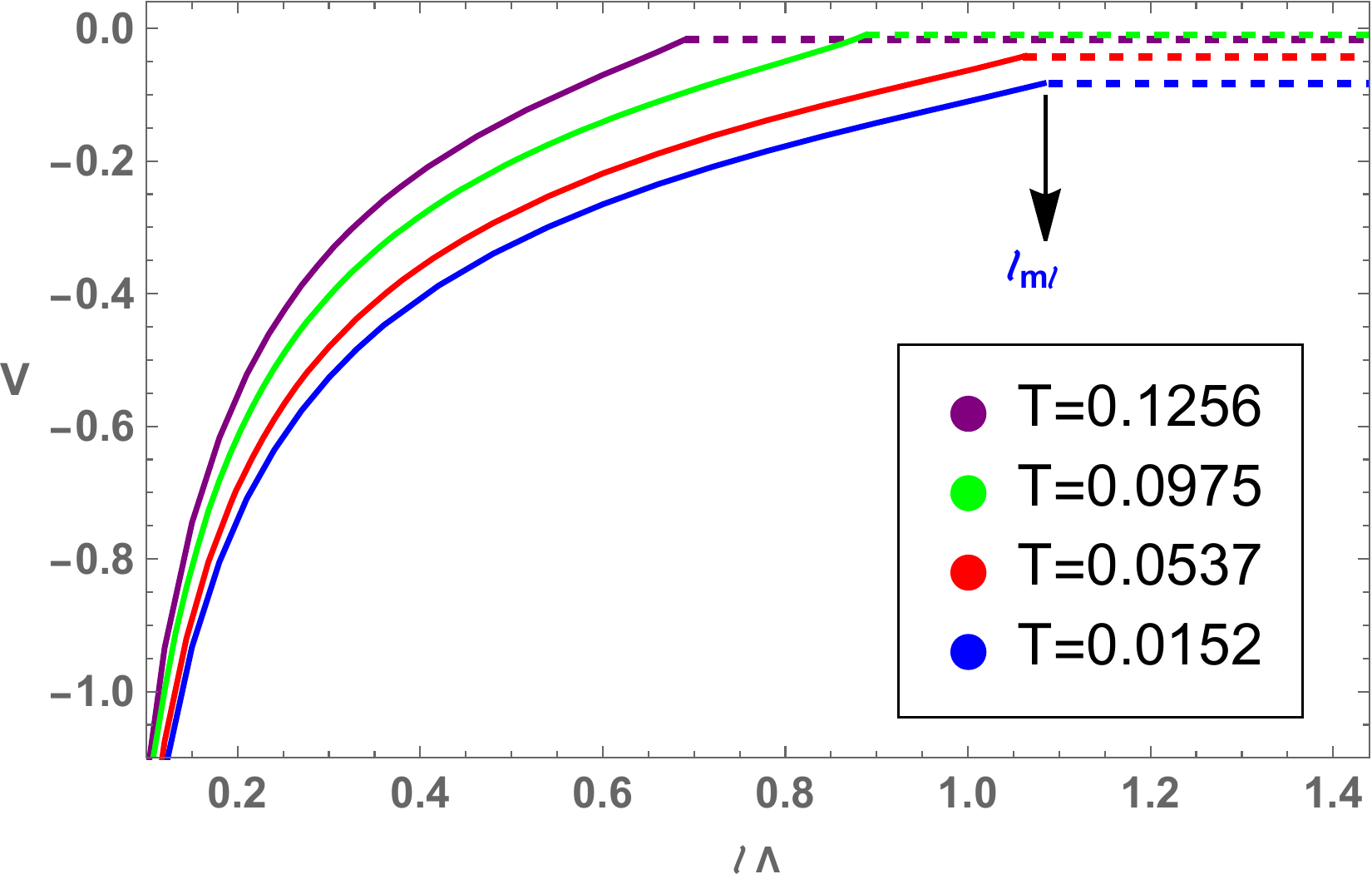}
\includegraphics[width=59mm]{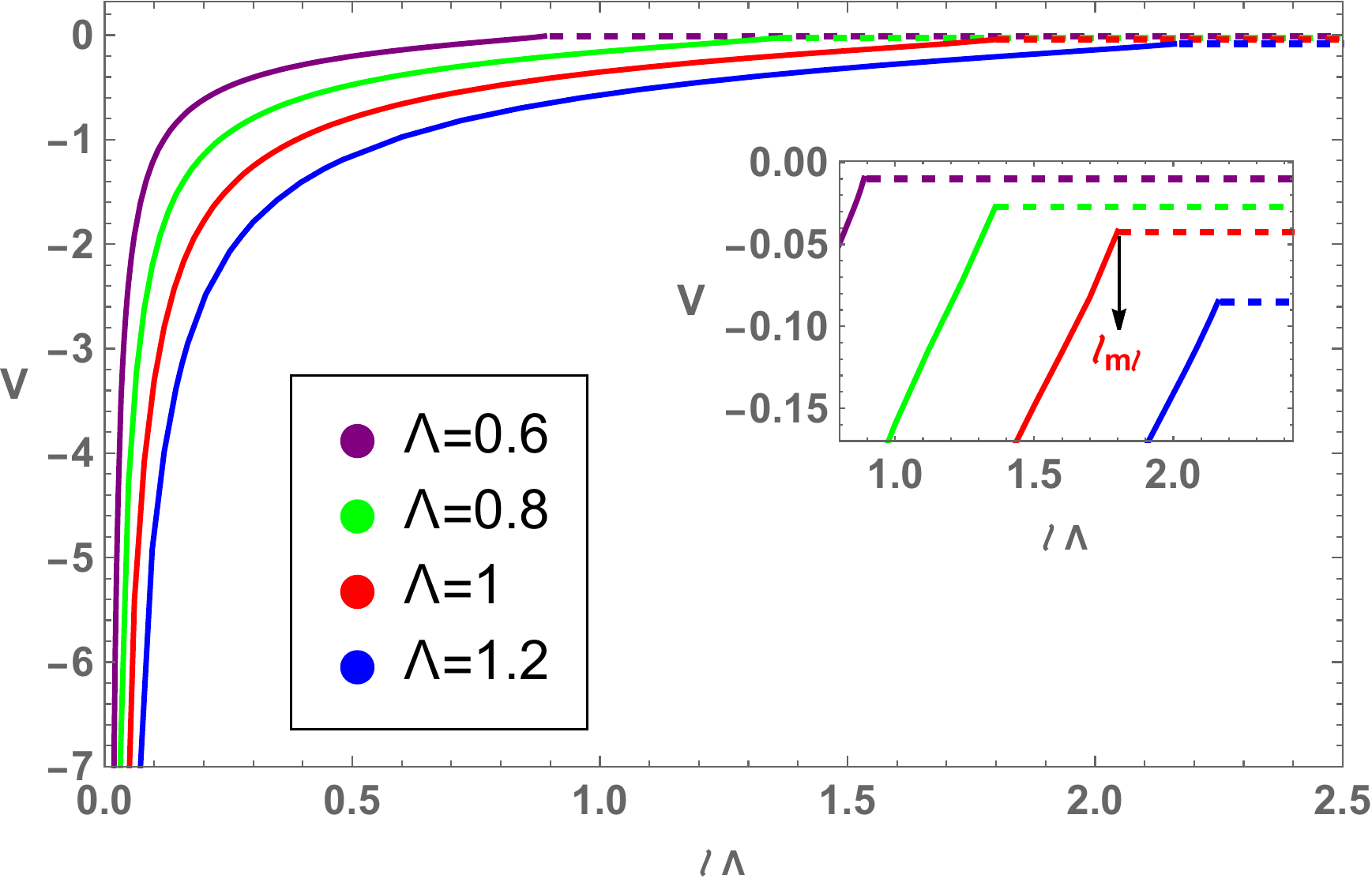}
\caption{Left: Potential energy of meson as a function of $\frac{\Lambda}{T}$ for different values of energy scale $\Lambda=1$ (blue), $\Lambda=0.7$ (green), $\Lambda=0.5$ (red) and $\Lambda=0.2$ (purple). We fix the values of the given model parameter $\phi_M=10$ and interquark distance $l=0.2$.
Middle: Potential energy of meson as a function of $l \Lambda$ for different values of the temperature $T=0.0152$ (blue), $T=0.0537$ (red), $T=0.0975$ (green) and  $T=0.1256$ (purple). We fix the values of the given model parameter $\phi_M=10$ and energy scale $\Lambda=0.6$.
Right: Potential energy of meson as a function of $l \Lambda$ for different values of energy scale $\Lambda=1.2$ (blue), $\Lambda=1$ (red), $\Lambda=0.8$ (green) and $\Lambda=0.6$ (purple). We fix the values of the given model parameter $\phi_M=10$ and the temperature $T=0.0975$.
}\label{f6}
\end{center}
\end{figure*}

\begin{itemize}
 \item[$-$] High temperature region ($\frac{\Lambda}{T} \ll 1$): The important point is that the conformal symmetry restores and the effect of $\Lambda$ that breaks conformal symmetry disappears. 
 Therefore, all the curves with different $\Lambda$ coincide. 
In the gravity side, the non-conformal background approaches $AdS_5$. 
 
  \item[$-$] Low temperature region ($\frac{\Lambda}{T} \gg 1$): The important feature is that  we can see the effect of non-conformality appears and the curves separate from each other.
 This is due to the fact that in this region effect of the energy scale $\Lambda$ dominates in comparison to the temperature T. At very low temperature region we see that the potential energy of meson gets approximately constant negative value. Note also that, by comparing the low temperature region and the middle panel of figure (\ref{f1}) in small $l$ we conclude that the physical effect of $\Lambda$ is the same, i.e. increasing $\Lambda$ leads to the more bounded quark-antiquark bound state.
  
\end{itemize}

\item Middle panel

It is observed that at finite temperature case the potential energy of meson is always negative in agreement with the left panel.
Another point is that, at fixed temperature $T$ and the energy scale $\Lambda$ increasing the interquark distance $l$ leads to the less bound state of meson. We can also observed that the potential energy of the pair increases monotonically and reaches to the maximum value corresponding to the special length, \textit{melting length}, $l_{m \ell}$. Beyond $l_{m \ell}$, it is clearly seen that $\cup$-shape open string configuration of the pair in the gravity side will break into two disconnected straight shape of open strings. Physical interpretation in the gauge theory side is that for $l<l_{m \ell}$ we have meson bound state, while for $l>l_{m\ell}$ the meson melting occurs in the plasma. Consequently, this indicates that for $l>l_{m\ell}$ the potential energy is independent of $l$ suggesting $q$, $\bar{q}$ pair will be broken. 
It is noticed that the connected string configuration that is appropriated for $l<l_{m\ell}$ and describes the potential energy of bound state of a quark and an anti-quark is denoted by a solid line, while the disconnected string configuration that is appropriated for $l>l_{m\ell}$ is denoted by a flat dashed line. 
Another feature is that, the melting length $l_{m\ell}$ is different for different temperatures. Since, the effect of temperature $T$ and interquark distance $l$ are opposite to each other then it is anticipated that $l_{m\ell}$ occurs at larger $l$ for lower $T$. 
In other words, for given values of model parameter $\phi_M$ and energy scale $\Lambda$ we have $l_{m\ell}^{T=0.0152}>l_{m\ell}^{T=0.0537}>l_{m\ell}^{T=0.0975}>l_{m\ell}^{T=0.1256}$. In addition, when we probe the theory at very small $l$ all the curves coincide, since the non-conformal field theory becomes conformal in the UV limit.
\item Right panel

As anticipated, at small $l$ (UV limit)  where we reach to the fixed point of the theory all the curves coincide. This indicates that in this limit the theory becomes effectively conformal. Also, in this panel similar to the middle panel the potential energy of the quark-antiquark pair increases monotonically and reaches to the maximum value.
At intermidiate $l$ the value of meson potential energy depends on $\Lambda$. It is clearly seen by fixing $l$ and $T$ increasing $\Lambda$ decreases the potential energy and the meson gets more bounded that is in agreement with the left panel. In addition, $l_{m\ell}$ changes by varying  $\Lambda$ and when we increase $\Lambda$ we see that $l_{m\ell}$ occurs at larger values of interquark distance $l$ and lower potential energy. In other words, for given $\phi_M$ and $T$ we have $l_{m\ell}^{\Lambda=1.2}>l_{m\ell}^{\Lambda=1}>l_{m\ell}^{\Lambda=0.8}>l_{m\ell}^{\Lambda=0.6}$. 
It can be described with the point that when we increase $\Lambda$, the meson gets more stable bound state as mentioned above. Therefore, it is reasonable that when we increase $\Lambda$, the meson dissociates at larger values of interquark distance $l$. Note also that, from the middle and right panels we see that the effect of $T$ and $\Lambda$ on $l_{m\ell}$ are opposite to each other.

\end{itemize}
\begin{figure*}[ht]
\begin{center}
\includegraphics[width=73mm]{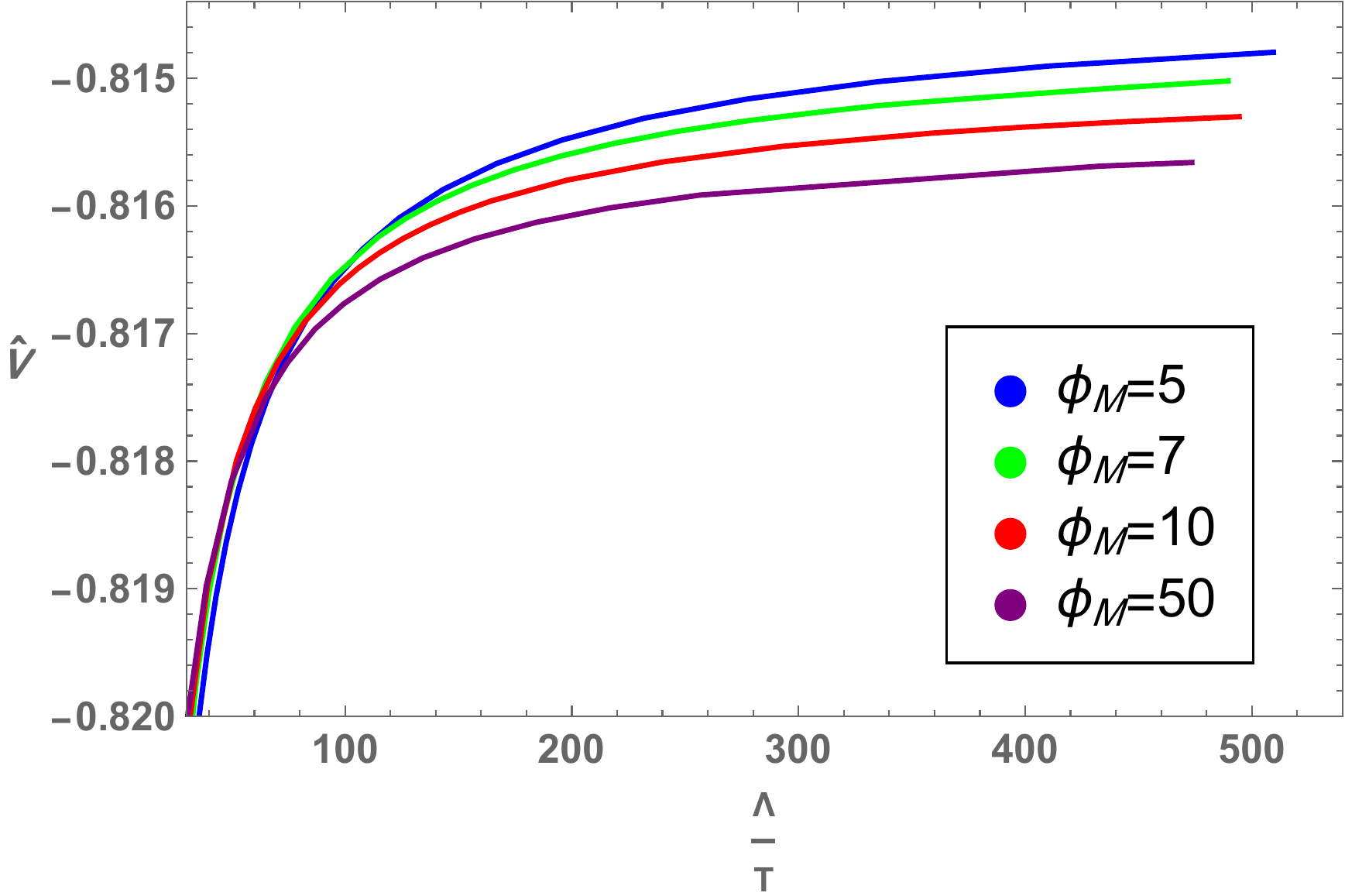}
\caption{
 Relative potential energy of meson $\hat{V}=\frac{V-V_{AdS}(T=0.0438)}{V_{AdS}(T=0.0438)}$ as a function of $\frac{\Lambda}{T}$ for different values of the model parameter $\phi_M=50$ (purple), $\phi_M=10$ (red), $\phi_M=7$ (green) and $\phi_M=5$ (blue).  We fix the values of the energy scale $\Lambda=0.4$ and interquark distance $l=0.35$.
}\label{f8}
\end{center}
\end{figure*}
In figure (\ref{f8}), we depict the relative potential energy of meson $\hat{V}=\frac{V-V_{AdS}(T=0.0438)}{V_{AdS}(T=0.0438)}$ at finite temperature case for fixed values of the energy scale $\Lambda=0.4$ and interquark distance $l=0.35$ and different values of the given model parameter $\phi_M=50$ (purple), $\phi_M=10$ (red), $\phi_M=7$ (green) and $\phi_M=5$ (blue). 
It is seen that the value of $\hat{V}$ increases monotonically by increasing the temperature.
 Also, when we are in the high temperature region ($\frac{\Lambda}{T} \ll 1$), the relative potential energy of meson $\hat{V}$ are equal for different values of $\phi_M=7$ and $\phi_M=20$. This is due to the fact that at high temperature the field theory is almost conformal and hence two curves coincide. 
On the other hand, at low temperature region ($\frac{\Lambda}{T} \gg 1$) the value of $\hat{V}$ depends on $\phi_M$. When we increase $\phi_M$, the value of $\hat{V}$ decreases and two curves deviate from each other.
Consequently, the meson potential energy can be considered as a measure of non-conformality of the theory at finite temperature case.  


\section{conclusion and remarks} \label{gr4}
In this paper we are interested in understanding in this research is to know the effect of energy scale $\Lambda$, that breaks the conformal symmetry, model parameter $\phi_M$, that controls the difference in degrees of freedom between UV and IR fixed points and interquark distance $l$ on the meson potential energy $V$ at both zero and finite temperature $T$.

Our main findings can be summarized as follows:
\begin{itemize}
\item 
Interestingly, we find that parameters of the Cornell potential, i.e. $\kappa$ and $C$ depends on the $\Lambda$ and $\Delta N$ while $\sigma_s$ just depends on the $\Lambda$. Therefore, in our model the Cornell potential as a function of $l$, $\Lambda$ and $\Delta N$ can be rewritten in the form of the  equation (\ref{cornel28}).
\item
The potential energy of meson increases by raising the interquark distance at zero and finite temperature. At zero temperature case we observe that there is a specific length $l_{\star}$ that beyond $l_{\star}$ the meson enjoys the linear regime of the Cornell potential. For small values of $l$ when we increase $\Lambda$, the meson potential energy decreases and the meson gets more bounded while at large $l$ this behavior reverses.
\item
By concentrating on the linear regime of the meson potential energy we find that making larger $\Lambda$,  both $l_{\star}$  and QCD string tension $\sigma_s$ increase monotonically. Therefore, at larger $\Lambda$ the meson experiences the linear regime at larger $l$.

\item
By raising the temperature of the system the meson potential energy increases. Also, at fixed $T$ when we increase $\Lambda$ the meson potential decreases that is the meson gets more bounded. Therefore, the effect of $T$ and $\Lambda$ are opposite to each other. 
\item
By investigating the meson potential energy as a function of $\frac{\Lambda}{T}$ we observe that at very high temperature region ($\frac{\Lambda}{T} \ll 1$) all curves corresponding with different values of $\Lambda$ or $\phi_M$ coincide because of restoring conformal symmetry. Although, at very low temperature region ($\frac{\Lambda}{T} \gg 1$) all curves deviate from each other. Also, at fixed low temperature region when we increase $\phi_M$ the meson potential energy increases that is the meson gets more bounded.
\item
 We observe that at finite temperature there is a \textit{melting length} $l_{m\ell}$ where beyond $l_{m\ell}$ the meson bound state dissociates in the plasma.
 Another point is that $l_{m\ell}$ is different for different temperatures of the system, i.e. $l_{m\ell}^{T=lower}>l_{m\ell}^{T=higher}$ and  occurs at different potentials.
Although, for different $\Lambda$ we have  $l_{m\ell}^{\Lambda=larger}>l_{m\ell}^{\Lambda=smaller}$. Also, our results show that for smaller $\Lambda$ the value of $l_{m\ell}$ occurs at higher potential.
\end{itemize}

By noting the numerical results for meson potential energy at zero and finite temperature cases the connection between them can be considered. A very interesting observation is that from the right panel of figure (\ref{f1}) and figure (\ref{f8}), i.e. at both zero and finite temperature cases, the relative meson potential energy can be considered as a good measure of non-conformality of the theory. Furthermore, at both cases, i.e. from the middle panel of figure (\ref{f1}) and the right panel of figure (\ref{f6}) it is observed that when we consider small $l$, increasing $\Lambda$ decreases meson potential energy that is the meson state gets more bounded. Also, at both cases working at \textit{very} small $l$ (UV limit) all the curves coincide. This is due to the fact that conformal symmetry restores at both cases at very small $l$.
In addition, from the middle panel of figure (\ref{f1}) and the left panel of figure (\ref{f6}) we observe that at very low temperature and small $l$ by increasing the $\Lambda$ meson potential energy decreases and the meson gets more bounded. Totally, our numerical results show that at finite temperature case for any choice of parameters increasing $\Lambda$ leads to the more stable bound state of meson while at zero temperature case just for small and large $l$ this effect can be observed.

It is important to note that the study of heavy quark-antiquark potential energy by trying to reconstruct the metric via solving the equation of motion and to fit the Cornell potential in a precise way is considered in different models. In \cite{Pirner:2009gr} by introducing the parameter Lambda in the warp factor with some logarithmic correction they try to reconstruct the metric and fix the parameter Lambda in such a way that leads to the good agreement with the Coulomb part and Linear part of the Cornell potential. In addition, utilizing this modified metric the heavy quark-antiquark potential is calculated for all length scales numerically.
Note also that in \cite{He:2010ye} the holographic QCD model is constructed such that the deformed warp factor is introduced in three cases.
In the case with only logarithmic correction that the parameter Lambda is introduced in the warp factor, the heavy quark-antiquark potential can be fitted perfectly. 
In the case with both logarithmic and quadratic corrections, the Coulomb part is in good agreement with the Cornell potential, the linear part is parallel to the Cornell potential, however, the value of the quark-antiquark potential is a little bit higher than the experimental data.
In the case with only quadratic correction, the resulted heavy quark-antiquark potential has both Coulomb part and linear part and fits the Cornell potential qualitatively well. 
In fact, the model with only logarithmic correction in the deformed warp factor can fit the heavy quark potential better than the model with only quadratic correction.
However, the results of \cite{He:2010ye} show that in the case with only quadratic correction, the Andreev-Zakharov model \cite{Andreev:2006ct} is favored to fit the heavy quark-antiquark potential.
For more details see  \cite{Andreev:2006ct,Pirner:2009gr,He:2010ye,Li:2011hp,Cai:2012xh} and references therein.

We would like to emphasize that, in our model we try to find out the parameters of the Cornell potential numerically. In fact, we produced the heavy quark-antiquark potential at zero and finite temperature cases for different parameters of the theory qualitatively.  In addition, we investigate the effect of the $\Lambda$ and $\phi_M$ on the meson potential  and the dependence of parameters of the Cornell potential, i.e. $\kappa$,  $\sigma_s$ and $C$ to the $\Lambda$ and $\Delta N$ at zero temperature case. Also, the effect of the $\Lambda$, $\phi_M$ and $T$ on the meson potential is studied at finite temperature case. Our approach in this research is completely numeric and our results are qualitative.

\section*{Acknowledgement}
We would like to thank kindly for helpful comments and discussions with M. Ali-Akbari, M. Lezgi and M. Rahimi.



\begin{thebibliography}{99} 
\bibitem{Maldacena:1997re}
J.~M.~Maldacena,
``The Large N limit of superconformal field theories and supergravity,''
Int. J. Theor. Phys. \textbf{38}, 1113-1133 (1999)
[arXiv:hep-th/9711200 [hep-th]].

\bibitem{Witten:1998qj}
E.~Witten,
``Anti-de Sitter space and holography,''
Adv. Theor. Math. Phys. \textbf{2}, 253-291 (1998)
[arXiv:hep-th/9802150 [hep-th]].

\bibitem{Gubser:1998bc}
S.~S.~Gubser, I.~R.~Klebanov and A.~M.~Polyakov,
``Gauge theory correlators from noncritical string theory,''
Phys. Lett. B \textbf{428}, 105-114 (1998)
[arXiv:hep-th/9802109 [hep-th]].

\bibitem{CasalderreySolana:2011us}
J.~Casalderrey-Solana, H.~Liu, D.~Mateos, K.~Rajagopal and U.~A.~Wiedemann,
``Gauge/String Duality, Hot QCD and Heavy Ion Collisions,''
[arXiv:1101.0618 [hep-th]].

\bibitem{Ammon:2015wua}
M.~Ammon and J.~Erdmenger,
``Gauge/gravity duality: Foundations and applications,''
Cambridge Univ. Pr., Cambridge, UK, 2015.

\bibitem{Kruczenski:2003be}
M.~Kruczenski, D.~Mateos, R.~C.~Myers and D.~J.~Winters,
``Meson spectroscopy in AdS / CFT with flavor,''
JHEP \textbf{07}, 049 (2003)
[arXiv:hep-th/0304032 [hep-th]].

\bibitem{Kruczenski:2003uq}
M.~Kruczenski, D.~Mateos, R.~C.~Myers and D.~J.~Winters,
``Towards a holographic dual of large N(c) QCD,''
JHEP \textbf{05}, 041 (2004)
[arXiv:hep-th/0311270 [hep-th]].

\bibitem{Kobayashi:2006sb}
S.~Kobayashi, D.~Mateos, S.~Matsuura, R.~C.~Myers and R.~M.~Thomson,
``Holographic phase transitions at finite baryon density,''
JHEP \textbf{02}, 016 (2007)
[arXiv:hep-th/0611099 [hep-th]].

\bibitem{Babington:2003vm}
J.~Babington, J.~Erdmenger, N.~J.~Evans, Z.~Guralnik and I.~Kirsch,
``Chiral symmetry breaking and pions in nonsupersymmetric gauge / gravity duals,''
Phys. Rev. D \textbf{69}, 066007 (2004)
[arXiv:hep-th/0306018 [hep-th]].

\bibitem{Sakai:2004cn}
T.~Sakai and S.~Sugimoto,
``Low energy hadron physics in holographic QCD,''
Prog. Theor. Phys. \textbf{113}, 843-882 (2005)
[arXiv:hep-th/0412141 [hep-th]].

\bibitem{Sakai:2005yt}
T.~Sakai and S.~Sugimoto,
``More on a holographic dual of QCD,''
Prog. Theor. Phys. \textbf{114}, 1083-1118 (2005)
[arXiv:hep-th/0507073 [hep-th]].

\bibitem{Erlich:2005qh}
J.~Erlich, E.~Katz, D.~T.~Son and M.~A.~Stephanov,
``QCD and a holographic model of hadrons,''
Phys. Rev. Lett. \textbf{95}, 261602 (2005)
[arXiv:hep-ph/0501128 [hep-ph]].

\bibitem{Karch:2006pv}
A.~Karch, E.~Katz, D.~T.~Son and M.~A.~Stephanov,
``Linear confinement and AdS/QCD,''
Phys. Rev. D \textbf{74}, 015005 (2006)
[arXiv:hep-ph/0602229 [hep-ph]].

\bibitem{Li:2011hp}
D.~Li, S.~He, M.~Huang and Q.~S.~Yan,
``Thermodynamics of deformed AdS$_5$ model with a positive/negative quadratic correction in graviton-dilaton system,''
JHEP \textbf{09}, 041 (2011)
[arXiv:1103.5389 [hep-th]].

\bibitem{DeWolfe:2010he}
O.~DeWolfe, S.~S.~Gubser and C.~Rosen,
``A holographic critical point,''
Phys. Rev. D \textbf{83}, 086005 (2011)
[arXiv:1012.1864 [hep-th]].

\bibitem{DeWolfe:2011ts}
O.~DeWolfe, S.~S.~Gubser and C.~Rosen,
``Dynamic critical phenomena at a holographic critical point,''
Phys. Rev. D \textbf{84}, 126014 (2011)
[arXiv:1108.2029 [hep-th]].

\bibitem{Hajilou:2021wmz}
A.~Hajilou,
``Meson Excitation Time as a Probe of Holographic Critical Point,''
[arXiv:2111.09010 [hep-th]].


\bibitem{Amiri-Sharifi:2016uso}
S.~Amiri-Sharifi, M.~Ali-Akbari, A.~Kishani-Farahani and N.~Shafie,
``Double Relaxation via AdS/CFT,''
Nucl. Phys. B \textbf{909}, 778-795 (2016)
[arXiv:1601.04281 [hep-th]].

\bibitem{Ali-Akbari:2015bha}
M.~Ali-Akbari, F.~Charmchi, A.~Davody, H.~Ebrahim and L.~Shahkarami,
``Time-dependent meson melting in an external magnetic field,''
Phys. Rev. D \textbf{91}, 106008 (2015)
[arXiv:1503.04439 [hep-th]].

\bibitem{Bohra:2020qom}
H.~Bohra, D.~Dudal, A.~Hajilou and S.~Mahapatra,
``Chiral transition in the probe approximation from an Einstein-Maxwell-dilaton gravity model,''
Phys. Rev. D \textbf{103}, no.8, 086021 (2021)
[arXiv:2010.04578 [hep-th]].

\bibitem{Dudal:2021jav}
D.~Dudal, A.~Hajilou and S.~Mahapatra,
``A quenched 2-flavour Einstein\textendash{}Maxwell\textendash{}Dilaton gauge-gravity model,''
Eur. Phys. J. A \textbf{57}, no.4, 142 (2021)
[arXiv:2103.01185 [hep-th]].

\bibitem{AliAkbari:2012vt}
M.~Ali-Akbari and H.~Ebrahim,
``Thermalization in External Magnetic Field,''
JHEP \textbf{03}, 045 (2013)
[arXiv:1211.1637 [hep-th]].

\bibitem{Ali-Akbari:2013txa}
M.~Ali-Akbari and H.~Ebrahim,
``Chiral symmetry breaking: To probe anisotropy and magnetic field in quark-gluon plasma,''
Phys. Rev. D \textbf{89}, no.6, 065029 (2014)
[arXiv:1309.4715 [hep-th]].

\bibitem{Fang:2015ytf}
Z.~Fang, S.~He and D.~Li,
``Chiral and Deconfining Phase Transitions from Holographic QCD Study,''
Nucl. Phys. B \textbf{907}, 187-207 (2016)
[arXiv:1512.04062 [hep-ph]].

\bibitem{Callebaut:2011ab}
N.~Callebaut, D.~Dudal and H.~Verschelde,
``Holographic rho mesons in an external magnetic field,''
JHEP \textbf{03}, 033 (2013)
[arXiv:1105.2217 [hep-th]].

\bibitem{Arefeva:2020vae}
I.~Y.~Aref'eva, K.~Rannu and P.~Slepov,
``Holographic model for heavy quarks in anisotropic hot dense QGP with external magnetic field,''
JHEP \textbf{07}, 161 (2021)
[arXiv:2011.07023 [hep-th]].


\bibitem{Braga:2017bml}
N.~R.~F.~Braga, L.~F.~Ferreira and A.~Vega,
``Holographic model for charmonium dissociation,''
Phys. Lett. B \textbf{774}, 476-481 (2017)
[arXiv:1709.05326 [hep-ph]].

\bibitem{Abt:2019tas}
R.~Abt, J.~Erdmenger, N.~Evans and K.~S.~Rigatos,
``Light composite fermions from holography,''
JHEP \textbf{11}, 160 (2019)
[arXiv:1907.09489 [hep-th]].

\bibitem{Nakas:2020hyo}
T.~Nakas and K.~S.~Rigatos,
``Fermions and baryons as open-string states from brane junctions,''
JHEP \textbf{12}, 157 (2020)
[arXiv:2010.00025 [hep-th]].


\bibitem{Shuryak:2003xe}
E.~Shuryak,
``Why does the quark gluon plasma at RHIC behave as a nearly ideal fluid?,''
Prog. Part. Nucl. Phys. \textbf{53}, 273-303 (2004)
[arXiv:hep-ph/0312227 [hep-ph]].

\bibitem{Shuryak:2004cy}
E.~V.~Shuryak,
``What RHIC experiments and theory tell us about properties of quark-gluon plasma?,''
Nucl. Phys. A \textbf{750}, 64-83 (2005)
[arXiv:hep-ph/0405066 [hep-ph]].

\bibitem{Schwartz:2014sze}
M.~D.~Schwartz,
``Quantum Field Theory and the Standard Model,''

\bibitem{Wilson:1974sk}
K.~G.~Wilson,
``Confinement of Quarks,''
Phys. Rev. D \textbf{10}, 2445-2459 (1974)

\bibitem{Maldacena:1998im}
J.~M.~Maldacena,
``Wilson loops in large N field theories,''
Phys. Rev. Lett. \textbf{80}, 4859-4862 (1998)
[arXiv:hep-th/9803002 [hep-th]].

\bibitem{Brandhuber:1998bs}
A.~Brandhuber, N.~Itzhaki, J.~Sonnenschein and S.~Yankielowicz,
``Wilson loops in the large N limit at finite temperature,''
Phys. Lett. B \textbf{434}, 36-40 (1998)
[arXiv:hep-th/9803137 [hep-th]].





\bibitem{Andreev:2006ct}
O.~Andreev and V.~I.~Zakharov,
``Heavy-quark potentials and AdS/QCD,''
Phys. Rev. D \textbf{74}, 025023 (2006)
[arXiv:hep-ph/0604204 [hep-ph]].



\bibitem{Yang:2015aia}
Y.~Yang and P.~H.~Yuan,
``Confinement-deconfinement phase transition for heavy quarks in a soft wall holographic QCD model,''
JHEP \textbf{12}, 161 (2015)
[arXiv:1506.05930 [hep-th]].

\bibitem{Ali-Akbari:2015ooa}
M.~Ali-Akbari, F.~Charmchi, A.~Davody, H.~Ebrahim and L.~Shahkarami,
``Evolution of Wilson loop in time-dependent N=4 super Yang-Mills plasma,''
Phys. Rev. D \textbf{93}, no.8, 086005 (2016)
[arXiv:1510.00212 [hep-th]].

\bibitem{Hajilou:2017sxf}
A.~Hajilou, M.~Ali-Akbari and F.~Charmchi,
``A Classical String in Lifshitz\textendash{}Vaidya Geometry,''
Eur. Phys. J. C \textbf{78}, no.5, 424 (2018)
[arXiv:1707.00967 [hep-th]].

\bibitem{Ishii:2014paa}
T.~Ishii, S.~Kinoshita, K.~Murata and N.~Tanahashi,
``Dynamical Meson Melting in Holography,''
JHEP \textbf{04}, 099 (2014)
[arXiv:1401.5106 [hep-th]].

\bibitem{Hajilou:2018dcb}
A.~Hajilou and M.~Ali-Akbari,
``Meson Excitation at Finite Chemical Potential,''
Eur. Phys. J. C \textbf{79}, no.3, 254 (2019)
[arXiv:1804.07965 [hep-th]].

\bibitem{Buchel:2015saa}
A.~Buchel, M.~P.~Heller and R.~C.~Myers,
``Equilibration rates in a strongly coupled nonconformal quark-gluon plasma,''
Phys. Rev. Lett. \textbf{114}, no.25, 251601 (2015)
[arXiv:1503.07114 [hep-th]].

\bibitem{Lezgi:2020bkc}
M.~Lezgi, M.~Ali-Akbari and M.~Asadi,
``Nonconformality, subregion complexity, and meson binding,''
Phys. Rev. D \textbf{104}, no.2, 026001 (2021)
[arXiv:2011.11625 [hep-th]].

\bibitem{Janik:2015waa}
R.~A.~Janik, G.~Plewa, H.~Soltanpanahi and M.~Spalinski,
``Linearized nonequilibrium dynamics in nonconformal plasma,''
Phys. Rev. D \textbf{91}, no.12, 126013 (2015)
[arXiv:1503.07149 [hep-th]].

\bibitem{Gursoy:2015nza}
U.~Gursoy, M.~Jarvinen and G.~Policastro,
``Late time behavior of non-conformal plasmas,''
JHEP \textbf{01}, 134 (2016)
[arXiv:1507.08628 [hep-th]].


\bibitem{Benincasa:2005iv}
P.~Benincasa, A.~Buchel and A.~O.~Starinets,
``Sound waves in strongly coupled non-conformal gauge theory plasma,''
Nucl. Phys. B \textbf{733}, 160-187 (2006)
[arXiv:hep-th/0507026 [hep-th]].


\bibitem{Attems:2016ugt}
M.~Attems, J.~Casalderrey-Solana, D.~Mateos, I.~Papadimitriou, D.~Santos-Olivan, C.~F.~Sopuerta, M.~Triana and M.~Zilhao,
``Thermodynamics, transport and relaxation in non-conformal theories,''
JHEP \textbf{10}, 155 (2016)
[arXiv:1603.01254 [hep-th]].

\bibitem{Rahimi:2016bbv}
M.~Rahimi, M.~Ali-Akbari and M.~Lezgi,
``Entanglement entropy in a non-conformal background,''
Phys. Lett. B \textbf{771}, 583-587 (2017)
[arXiv:1610.01835 [hep-th]].

\bibitem{Ali-Akbari:2019zkf}
M.~Ali-Akbari, M.~Rahimi and M.~Asadi,
``Holographic mutual and tripartite information in a non-conformal background,''
Nucl. Phys. B \textbf{964}, 115329 (2021)
[arXiv:1907.08917 [hep-th]].

\bibitem{Asadi:2020gzl}
M.~Asadi,
``On volume subregion complexity in non-conformal theories,''
Eur. Phys. J. C \textbf{80}, no.7, 681 (2020)
[arXiv:2004.11306 [hep-th]].


\bibitem{Critelli:2016cvq}
R.~Critelli, R.~Rougemont, S.~I.~Finazzo and J.~Noronha,
``Polyakov loop and heavy quark entropy in strong magnetic fields from holographic black hole engineering,''
Phys. Rev. D \textbf{94}, no.12, 125019 (2016)
[arXiv:1606.09484 [hep-ph]].

\bibitem{Bohra:2019ebj}
H.~Bohra, D.~Dudal, A.~Hajilou and S.~Mahapatra,
``Anisotropic string tensions and inversely magnetic catalyzed deconfinement from a dynamical AdS/QCD model,''
Phys. Lett. B \textbf{801}, 135184 (2020)
[arXiv:1907.01852 [hep-th]].

\bibitem{He:2013qq}
S.~He, S.~Y.~Wu, Y.~Yang and P.~H.~Yuan,
``Phase Structure in a Dynamical Soft-Wall Holographic QCD Model,''
JHEP \textbf{04}, 093 (2013)
[arXiv:1301.0385 [hep-th]].

\bibitem{Gursoy:2007cb}
U.~Gursoy and E.~Kiritsis,
``Exploring improved holographic theories for QCD: Part I,''
JHEP \textbf{02}, 032 (2008)
[arXiv:0707.1324 [hep-th]].

\bibitem{Gursoy:2007er}
U.~Gursoy, E.~Kiritsis and F.~Nitti,
``Exploring improved holographic theories for QCD: Part II,''
JHEP \textbf{02}, 019 (2008)
[arXiv:0707.1349 [hep-th]].

\bibitem{Charmousis:2010zz}
C.~Charmousis, B.~Gouteraux, B.~S.~Kim, E.~Kiritsis and R.~Meyer,
``Effective Holographic Theories for low-temperature condensed matter systems,''
JHEP \textbf{11}, 151 (2010)
[arXiv:1005.4690 [hep-th]].

\bibitem{Dudal:2017max}
D.~Dudal and S.~Mahapatra,
``Thermal entropy of a quark-antiquark pair above and below deconfinement from a dynamical holographic QCD model,''
Phys. Rev. D \textbf{96}, no.12, 126010 (2017)
[arXiv:1708.06995 [hep-th]].

\bibitem{Arefeva:2018hyo}
I.~Aref'eva and K.~Rannu,
``Holographic Anisotropic Background with Confinement-Deconfinement Phase Transition,''
JHEP \textbf{05}, 206 (2018)
[arXiv:1802.05652 [hep-th]].


\bibitem{Eichten:1978tg}
E.~Eichten, K.~Gottfried, T.~Kinoshita, K.~D.~Lane and T.~M.~Yan,
``Charmonium: The Model,''
Phys. Rev. D \textbf{17}, 3090 (1978).


\bibitem{Bruni:2018dqm}
R.~C.~L.~Bruni, E.~Folco Capossoli and H.~Boschi-Filho,
``Quark-antiquark potential from a deformed AdS/QCD,''
Adv. High Energy Phys. \textbf{2019}, 1901659 (2019)
[arXiv:1806.05720 [hep-th]].
 
  \bibitem{Pirner:2009gr}
H.~J.~Pirner and B.~Galow,
``Strong Equivalence of the AdS-Metric and the QCD Running Coupling,''
Phys. Lett. B \textbf{679}, 51-55 (2009)
[arXiv:0903.2701 [hep-ph]].
  
\bibitem{He:2010ye}
S.~He, M.~Huang and Q.~S.~Yan,
``Logarithmic correction in the deformed $\rm AdS_5$ model to produce the heavy quark potential and QCD beta function,''
Phys. Rev. D \textbf{83}, 045034 (2011)
[arXiv:1004.1880 [hep-ph]].
  
\bibitem{Li:2011hp}
D.~Li, S.~He, M.~Huang and Q.~S.~Yan,
``Thermodynamics of deformed AdS$_5$ model with a positive/negative quadratic correction in graviton-dilaton system,''
JHEP \textbf{09}, 041 (2011)
[arXiv:1103.5389 [hep-th]].
  
\bibitem{Cai:2012xh}
R.~G.~Cai, S.~He and D.~Li,
``A hQCD model and its phase diagram in Einstein-Maxwell-Dilaton system,''
JHEP \textbf{03}, 033 (2012)
[arXiv:1201.0820 [hep-th]].
  
\end{thebibliography}
\end{document}